\documentclass[onecolumn, 11pt]{IEEEtran}
\usepackage{graphicx}
\usepackage{mathptmx}
\usepackage{amsmath}
\usepackage{amssymb}
\usepackage{theorem}
\usepackage{array}
\usepackage{url}

\def\BibTeX{{\rm B\kern-.05em{\sc i\kern-.02
5em b}\kern-.08em\kern-.1667em\lower.7ex\hbox{E}\kern-.125emX}}

\newcommand{\linespacing}{1}
\renewcommand{\baselinestretch}{\linespacing}

\usepackage{graphicx}

\ifCLASSINFOpdf
\else
\fi
\begin{document}
%\title{Using the Style File IEEEtran.sty}
\title{Performance bounds and codes design criteria for channel decoding with a-priori information}

\author{\IEEEauthorblockN{Andrea Abrardo}\\
\IEEEauthorblockA{Department of Information Engineering\\
Via Roma 56, 53100 Siena, Italy\\
email: abrardo@dii.unisi.it} }


%Telephone: (800) 555--1212, Fax: (888) 555--1212}


\maketitle

\begin{abstract}
%\boldmath
In this article we focus on the problem of channel decoding in
presence of a-priori information. In particular, assuming that the
a-priori information reliability is not perfectly estimated at the
receiver, we derive a novel analytical framework for evaluating the
decoder's performance. It is derived the important result that a
"good code", i.e., a code which allows to fully exploit the
potential benefit of a-priori information, must associate
information sequences with high Hamming weights to codewords with
low Hamming weights. Basing on the proposed analysis, we analyze the
performance of convolutional codes, random codes, and turbo codes.
Moreover, we consider the transmission of correlated binary sources
from independent nodes, a problem which has several practical
applications, e.g. in the case of sensor networks. In this context,
we propose a very simple joint source-channel turbo decoding scheme
where each decoder works by exploiting a-priori information given by
the other decoder. In the case of block fading channels, it is shown
that the inherent correlation between information signals provide a
form of non-cooperative diversity, thus allowing joint
source-channel decoding to outperform separation-based schemes.

\end{abstract}

\section{Introduction}

In most digital applications source and channel coding are treated
as separate schemes, and the common approach of channel coding is to
consider source encoded streams as statistically independent
streams. However, in several situations it is not possible, or not
convenient, to let source coding eliminating all intrinsic data
redundancy. In this cases, the decoder can exploit such a residual
(or total) redundancy in its effort of combating noise by performing
\emph{joint source-channel decoding} (JSCD). However, one of the
main problem which arises in JSCD is represented by implementation
complexity of the decoder, which in general increases to take into
account the memory of the information source. As an example, for a
first-order Markov source which is protected by convolutional codes,
 the optimum JSCD scheme is the
maximum a posteriori (MAP) sequence decoder based on a
"super-trellis". The number of the "super-trellis" states is the
product of the number of states of the convolutional trellis and the
Markov trellis. Some methods have been proposed to reduce the number
of trellis states, which result in suboptimum MAP decoders based on
symbol or bit-level \cite{Hagenauer}, \cite{Park}, \cite{Lamy},
\cite{Jeanne}, \cite{Liu}, \cite{Crespo}. Suboptimal codes aim at
presenting redundancy of information sources as a-priori information
(API) at the input of channel decoder/demodulator, so that iterative
schemes can be easily derived where at each iteration API can be
easily enclosed in the decoder
 without substantially increasing the receiver complexity. In
 particular, when API is presented at bit-level,
 the use of channel decoding schemes can be easily extended to all MAP-based decoding
 schemes, e.g., turbo decoders and LDPC
 decoders \cite{Zhu}, \cite{Pu}.

Another field where JSCD is gaining its momentum is the transmission
of detected signals observed at different nodes in Wireless Sensor
Networks (WSNs) \cite{Akyildiz}. In the case of a single collector
node (the access point), the study of efficient transmission
mechanisms is often referred to as reach-back channel problem
\cite{Barros}, \cite{Gupta}, \cite{Gamal}. In an attempt to exploit
the intrinsic correlation among data, many works have recently
focussed on the design of source coding schemes that approach the
Slepian-Wolf fundamental limit on the achievable compression rates
\cite{Aaron}, \cite{Bajcsy}, \cite{Deslauriers}, \cite{Xiong}, thus
applying the separation principle. However, the design of good
practical source codes for correlated sources is still an open
problem. Besides, separation between source and channel coding may
lead to catastrophic error propagation. Eventually, the traditional
code design requires that the correlation between the two sources is
known in the encoding process, a requisite that in many applications
(e.g., when the nodes are randomly placed in an environment) can be
hardly achieved. In an attempt to overcome this impairment, several
papers have proposed JSCD schemes where the correlated sources are
channel encoded at a reduced rate (with respect to the uncorrelated
case). The reduced reliability due to channel coding rate reduction
can be compensated by exploiting correlation among different
information sources at the channel decoder \cite{Garcia-Frias1},
\cite{Garcia-Frias2}, \cite{Garcia-Frias3}, \cite{Mondin1},
\cite{Maramatsu}. In particular, exploiting correlation by means of
API has been shown to achieve very good performance.\\
Although the great attention that has been given to these topics in
the recent literature, the problem of designing good codes in
presence of API has not been addressed so far. This is because it is
generally assumed that good codes in the classical case (no API) are
still good in presence of API. In an attempt to fill this lack, in
this paper we derive some useful bounds for the bit error
probability which establish that the performance depends not only on
codewords' weights, as in traditional decoding, but also on
information data weights. The proposed analysis allows to give an
insight into the design of good codes, i.e., channel codes which
permit to take the best advantage from exploiting API at the
decoder. Furthermore, we consider the transmission of correlated
binary sources from independent nodes and we propose a very simple
JSCD scheme, where
each decoder works by exploiting API given by the other decoder.\\
This paper is structured as follows. In Section II, we derive the
pairwise error probability in presence of API at the decoder. In
Section III we validate the analysis in the uncoded case. In Section
IV we provide an analytical study for evaluating performance in
three different coded scenarios: (\emph{i}) convolutional codes,
(\emph{ii}) random codes with infinite length, and (\emph{iii})
turbo codes. Eventually, in Section IV we propose a JSCD scheme for
decoding correlated binary sources from independent nodes. Finally,
concluding remarks are given in Section IV.

\section{Pairwise error probability evaluation}

We consider an i.i.d binary source signal ${\mathbf{x}}$ of length
$k$ which is channel encoded with rate $r=k/n$ and denote by
${\mathbf{c}}$ the binary coded signal of length $n$. We assume that
a side-information $\tilde{x}_i = 0/1$ about the message
${\mathbf{x}}$ is available at the decoder and we denote to as
$\rho$ the side-information reliability, i.e., $\rho =
Pr\left(\tilde{x}_i = x_i\right)$. Let introduce the a-priori
log-likelihood terms $L({{x}}_i) = ln \left[
\frac{Pr\left({{x}}_i=0\right)}{Pr\left({{x}}_i=1\right)}\right]$
($ln$ represents the natural logarithm). Given these notations, it
is easy to derive $L(x_i) = L(x) \times (-2\tilde{x}_i+1)$, where
$L(x) = ln \left( \frac{\rho}{1-\rho}\right)$. Of course, in order
to fruitfully exploit the side information, the channel decoder must
generate an estimate of the reliability $\rho$. This can be easily
obtained by evaluating the number of zeros of the XOR between the
received sequences. In the following, we assume that an estimation
$\tilde{\rho}$ is available at the decoder. Accordingly, we
introduce $\tilde{L}(x) = ln \left(
\frac{\tilde{\rho}}{1-\tilde{\rho}}\right)$.\\
Let us denote by ${\mathbf{y}}({\mathbf{x}})$ the transmitted signal
and assume a binary antipodal modulation scheme, so that
${\mathbf{y}}({\mathbf{x}}) = -2{\mathbf{c}}({\mathbf{x}})+1$.
Eventually, assuming an AWGN channel model, we can express the
received signal ${\mathbf{z}}$ as:
\begin{equation}
    \begin{array}{cl}
    {\mathbf{z}} = \sqrt{2 r \xi_{b}} \times {\mathbf{y}}({\mathbf{x}}) + {\mathbf{\eta}}
    \end{array}
 \label{eq3AAnext1}
\end{equation}
where $\eta_{i}$ are Gaussian random noise terms with zero mean and
variance $N_0$ and $\xi_{b}$ is the energy per bit. \\
Denoting by $\tilde{{\mathbf{x}}}$ the side information at the
decoder, the MAP decoding rule can be expressed as:
\begin{equation}
    \begin{array}{cl}
    \hat{{\mathbf{x}}} = \mathop{arg ~
max}\limits_{{{\mathbf{x}}}}
~\Pr\left\{{\mathbf{x}}|\tilde{\rho},\tilde{{{\mathbf{x}}}},{{\mathbf{z}}}\right\}
    \end{array}
 \label{eq3AAprev6}
\end{equation}
By using the Bayes' rule and neglecting any constant term (i.e., the
terms which do not depend on ${{\mathbf{x}}}$), it is now
straightforward to get from (\ref{eq3AAprev6}) the equivalent
decoding rule:
\begin{equation}
    \begin{array}{cl}
    \hat{{\mathbf{x}}} = \mathop{arg ~
max}\limits_{{{\mathbf{x}}}}
~\Pr\left\{{{\mathbf{z}}}|{\mathbf{x}}\right\}Pr\left\{{{\mathbf{x}}}|\tilde{\rho},\tilde{{{\mathbf{x}}}}\right\}
    \end{array}
 \label{eq3AAprev6bis}
\end{equation}
Using the AWGN assumption and substituting for ${{\mathbf{z}}}$ the
expression given in (\ref{eq3AAnext1}) it is easy to derive:
%$Pr\left\{{{\mathbf{u}}_x}\right}
%\right\}$ does not depend on the information signals
%$\left\{{{\mathbf{x}}},{{\mathbf{y}}}\right\}$, we can rewrite the
%optimum decoding problem as:

%
%
% We are thus interested in
%sub-optimum joint decoding algorithms based on iterative approaches,
%where each decoder works on its own basing on a-priori information
%given by the other decoder. In particular, let assume without loss
%of generality that the decoder has come into the estimation
%$\tilde{{\mathbf{y}}}$ of the transmitted information signal
%${{\mathbf{y}}}$ and denote to as $q$ the probability of correct
%detection, i.e., $q = Pr\left\{\tilde{{{y}}}_i = y_i\right\}$.
%%Note that several decoding
%%algorithms may provide at their outputs such a probabilities, as for
%%example the Soft-Input Soft-output scheme proposed in
%%\cite{Benedetto}.

%Taking into account the correlation model introduced in the previous
%Section, it is now straightforward to derive the a-priori
%probability for the other decoder $p = Pr\left\{x_i =
%\tilde{{{y}}}_i\right\}$ as:
%\begin{equation}
%    \begin{array}{cl}
%p = \rho \times q + (1-\rho) (1-q)
%    \end{array}
% \label{eq3AAgt}
%\end{equation}
%The decoding problem can now be formulated as:
%\begin{equation}
%    \begin{array}{cl}
%    \tilde{{\mathbf{x}}} = \mathop{arg ~
%max}\limits_{{{\mathbf{x}}}}
%Pr\left\{{{\mathbf{x}}}|\tilde{{\mathbf{y}}},{{\mathbf{u}}_x}
%\right\}
%    \end{array}
% \label{eq3AA}
%\end{equation}
%which yields:
\begin{equation}
    \begin{array}{cl}
    \tilde{{\mathbf{x}}} = \mathop{arg ~
max}\limits_{{{\mathbf{x}}}} \left[\sqrt{2
r\xi_{b}}\sum\limits_{i=0}^{n-1} z_{i}y_i + N_0 \times
ln\left(Pr\left\{{{\mathbf{x}}}|\tilde{\rho},\tilde{{{\mathbf{x}}}}\right\}\right)\right]
    \end{array}
 \label{eq4AA}
\end{equation}
Let us now denote by ${{\mathbf{x}}_t}$ the transmitted information
signal, and by ${{\mathbf{x}}_e} \neq {{\mathbf{x}}_t}$ the
estimated sequence. Moreover, let denote by ${{\mathbf{y}}_e} \neq
{{\mathbf{y}}_t}$ the corresponding codewords. The pairwise error
probability conditioned to $\tilde{{{\mathbf{x}}}}$ can be defined
as the probability that the metric (\ref{eq4AA}) evaluated for
${\mathbf{y}} = {{\mathbf{y}}_e}$ and ${\mathbf{x}} =
{{\mathbf{x}}_e}$ is higher than that evaluated for ${\mathbf{y}} =
{{\mathbf{y}}_t}$ and ${\mathbf{x}} = {{\mathbf{x}}_t}$. Such a
probability can be expressed as:
\begin{equation}
    \begin{array}{cl}
    P_e\left({{\mathbf{x}}_t},{{\mathbf{x}}_e}|\tilde{{{\mathbf{x}}}}\right) =
    \Pr\left\{\sqrt{2 r
\xi_{b}}\sum\limits_{i=0}^{n-1} z_{i}\left(y_{i,e}-y_{i,t}\right)
 - N_0 \times
ln\left(\frac{Pr\left\{{{\mathbf{x}}_t}|\tilde{\rho},\tilde{{{\mathbf{x}}}}\right\}}{Pr\left\{{{\mathbf{x}}_e}|\tilde{\rho},\tilde{{{\mathbf{x}}}}\right\}}\right)
> 0\right\}
    \end{array}
 \label{eq5AA}
\end{equation}
Substituting for ${\mathbf{z}}$ in (\ref{eq5AA}) the expression
given in (\ref{eq3AAnext1}), it is straightforward to obtain:
\begin{equation}
    \begin{array}{cl}
    P_e\left({{\mathbf{x}}_t},{{\mathbf{x}}_e}|\tilde{{{\mathbf{x}}}}\right) =
    0.5 erfc\left[\sqrt{r d \gamma_{b}}  +\frac{1}{4\sqrt{r d \gamma_{b} }}ln\left(\frac{Pr\left\{{{\mathbf{x}}_t}|\tilde{\rho},\tilde{{{\mathbf{x}}}}\right\}}{Pr\left\{{{\mathbf{x}}_e}|\tilde{\rho},\tilde{{{\mathbf{x}}}}\right\}}\right)
    \right]
    \end{array}
 \label{eq6AA}
\end{equation}
where $\gamma_{b} = \frac{\xi_{b}}{N_0}$, $d =
D\left({{\mathbf{c}}_t},{{\mathbf{c}}_e}\right)$ is the Hamming
distance between ${{\mathbf{c}}_t}$ and ${{\mathbf{c}}_e}$ and
$erfc$ is the complementary error function.

To elaborate, we get from the hypothesis that ${{{\mathbf{x}}}}$ is
an i.i.d. sequence:
\begin{equation}
    \begin{array}{cl}
    \frac{Pr\left\{{{\mathbf{x}}_t}|\tilde{\rho},\tilde{{{\mathbf{x}}}}\right\}}{Pr\left\{{{\mathbf{x}}_e}|\tilde{\rho},\tilde{{{\mathbf{x}}}}\right\}}
    = \prod\limits_{i = 0}^{k-1} \frac{Pr\left\{{{{x}}_{i,t}}|\tilde{\rho},\tilde{{{{x}}}}_i\right\}}{Pr\left\{{{{x}}_{i,e}}|\tilde{\rho},\tilde{{{{x}}}}_i\right\}}
    \end{array}
 \label{eq6AA.0}
\end{equation}
Let us introduce the sequences ${\epsilon}_{i,t} = {x}_{i,t}
\bigoplus \tilde{x}_i$ and ${\epsilon}_{i,e} = {x}_{i,e} \bigoplus
\tilde{x}_i$, where $\bigoplus$ is the bit-wise XOR operator. By
exploiting the API $\tilde{{{{x}}}}_i$ and its estimated reliability
$\tilde{\rho}$, the $i$-th term in (\ref{eq6AA.0}) can be further
elaborated as:
\begin{equation}
    \begin{array}{cl}
    \frac{Pr\left\{{{{x}}_{i,t}}|\tilde{\rho},\tilde{{{{x}}}}_i\right\}}{Pr\left\{{{{x}}_{i,e}}|\tilde{\rho},\tilde{{{{x}}}}_i\right\}}
    = \frac{\tilde{\rho}^{{\epsilon}_{i,t}} \times \left(1-\tilde{\rho}\right)^{\bar{{\epsilon}}_{i,t}}}{\tilde{\rho}^{{\epsilon}_{i,e}} \times \left(1-\tilde{\rho}\right)^{\bar{{\epsilon}}_{i,e}}}
    = \left\{\begin{array}{ccccc}
              1 & if & {{{x}}_{i,t}} = {{{x}}_{i,e}} & ~ & ~ \\
              \frac{\tilde{\rho}}{1-\tilde{\rho}} & if & {{{x}}_{i,t}} \ne
              {{{x}}_{i,e}} & and & {{{\epsilon}}_{i,t}} =
              0 \\
              \frac{1-\tilde{\rho}}{\tilde{\rho}} & if & {{{x}}_{i,t}} \ne
              {{{x}}_{i,e}} & and & {{{\epsilon}}_{i,t}} = 1
            \end{array}
    \right.
    \end{array}
 \label{eq6.1}
\end{equation}
where $\bar{{\epsilon}}_{i,t}$ and $\bar{{\epsilon}}_{i,e}$ are the
NOT version of ${{\epsilon}}_{i,t}$ and ${{\epsilon}}_{i,e}$,
respectively. Hence, denoting by
$U\left({{\mathbf{x}}_t},{{\mathbf{x}}_e}\right)$ the set of indexes
such as ${{{x}}_{i,t}} \ne {{{x}}_{i,e}}$, i.e., ${{{x}}_{i,t}} \ne
{{{x}}_{i,e}}$ $\forall$ $i \in
U\left({{\mathbf{x}}_t},{{\mathbf{x}}_e}\right)$, we can write:
\begin{equation}
    \begin{array}{cl}
    \frac{Pr\left\{{{\mathbf{x}}_t}|\tilde{\rho},\tilde{{{\mathbf{x}}}}\right\}}{Pr\left\{{{\mathbf{x}}_e}|\tilde{\rho},\tilde{{{\mathbf{x}}}}\right\}}
    = \prod\limits_{i \in
    U\left({{\mathbf{x}}_t},{{\mathbf{x}}_e}\right)} \frac{Pr\left\{{{{x}}_{i,t}}|\tilde{\rho},\tilde{{{{x}}}}_i\right\}}{Pr\left\{{{{x}}_{i,e}}|\tilde{\rho},\tilde{{{{x}}}}_i\right\}}
    \end{array}
 \label{eq6.2}
\end{equation}
For the sake of notation clarity, we assume without loss of
generality that $U\left({{\mathbf{x}}_t},{{\mathbf{x}}_e}\right)$ is
the set $\{0,1,\ldots,w-1\}$, $w$ being the cardinality of
$U\left({{\mathbf{x}}_t},{{\mathbf{x}}_e}\right)$, i.e., $w =
D\left({{\mathbf{x}}_t},{{\mathbf{x}}_e}\right)$ is the Hamming
distance between ${{\mathbf{x}}_t}$ and ${{\mathbf{x}}_e}$. Hence,
we can write from (\ref{eq6.1}) and (\ref{eq6.2}):
\begin{equation}
    \begin{array}{cl}
    \frac{Pr\left\{{{\mathbf{x}}_t}|\tilde{\rho},\tilde{{{\mathbf{x}}}}\right\}}{Pr\left\{{{\mathbf{x}}_e}|\tilde{\rho},\tilde{{{\mathbf{x}}}}\right\}}
    =  \left(\frac{\tilde{\rho}}{1-\tilde{\rho}}\right)^{w-\sum\limits_{i = 0}^{w-1}
    {\epsilon}_{i,t}}
    \times \left(\frac{1-\tilde{\rho}}{\tilde{\rho}}\right)^{\sum\limits_{i = 0}^{w-1}{\epsilon}_{i,t}} = \left(\frac{\tilde{\rho}}{1-\tilde{\rho}}\right)^{w-\sum\limits_{i = 0}^{w-1}{2\epsilon}_{i,t}}  \end{array}
 \label{eq6.3}
\end{equation}
Denoting for the sake of simplicity ${\epsilon}_{i,t} =
{\epsilon}_{i}$, remembering that $\tilde{L}(x) = ln \left(
\frac{\tilde{\rho}}{1-\tilde{\rho}}\right)$, and introducing the
term $\tilde{w} = \sum\limits_{i=0}^{w-1} {\epsilon}_i$, it is now
straightforward to rewrite (\ref{eq6AA}) as:
\begin{equation}
    \begin{array}{cl}
    P_e\left({{\mathbf{x}}_t},{{\mathbf{x}}_e}|\tilde{{{\mathbf{x}}}}\right) =
    0.5erfc\left(\sqrt{{r d
\gamma_{b}}\left(1+\frac{1}{d}\frac{\tilde{L}(x)(w-2\tilde{w})}{4 r
\gamma_{b}}\right)^2}\right)
    \end{array}
 \label{eq6.3}
\end{equation}
It can be observed from (\ref{eq6.3}) that, if we condition to
$\tilde{w}$, the pairwise error probability depends on $d$ and $w$
rather than on the whole transmitted and estimated sequences
${{\mathbf{x}}_t}$ and ${{\mathbf{x}}_e}$. It is then possible to
write:
\begin{equation}
    \begin{array}{cl}
    P_e\left(d,w|\tilde{w}\right) =
    0.5erfc\left(\sqrt{{r d
\gamma_{b}}\left(1+\frac{1}{d}\frac{\tilde{L}(x)(w-2\tilde{w})}{4 r
\gamma_{b}}\right)^2}\right)
    \end{array}
 \label{eq6.4}
\end{equation}
Note that, according to the correlation model, ${\epsilon}_i$ are
i.i.d binary random term with $Pr\left\{{\epsilon}_i = 0\right\} =
\rho$ and $Pr\left\{{\epsilon}_i = 1\right\} = 1-\rho$. Hence,
$\tilde{w}$ is binomially distributed with parameters $w$ and
$1-\rho$, and the pairwise error probability can be eventually
derived as:
\begin{equation}
    \begin{array}{cl}
    P_e\left(d,w\right) =
    0.5\sum\limits_{\tilde{w}=0}^{w}erfc\left(\sqrt{{r d
\gamma_{b}}\left(1+\frac{1}{d}\frac{\tilde{L}(x)(w-2\tilde{w})}{4 r
\gamma_{b}}\right)^2}\right) {w \choose \tilde{w}}\rho^{w-\tilde{w}}
\times (1-\rho)^{\tilde{w}}
    \end{array}
 \label{eq6.5}
\end{equation}
The above expression is quite messy to manipulate. A significant
simplification occurs if we consider the following bound:
\begin{equation}
\begin{array}{c}
\left(1+\frac{1}{d}\frac{\tilde{L}(x)\left(w-2\tilde{w}\right)}{4 r
\gamma_{b}}\right)^2 \ge 1 +
\frac{2}{d}\frac{\tilde{L}(x)\left(w-2\tilde{w}\right)}{4 r
\gamma_{b}}
\end{array}
 \label{eqAA1}
\end{equation}
which is a tight lower bound for $r d \gamma_b
>> |\tilde{L}(x)\left(w-2\tilde{w}\right)|$ , i.e., when the error probability
is mainly determined by the codewords' distance rather than by the
beneficial effect of API.
%,i.e., for $\gamma_{b} >> \tilde{L}(x)$.
In this case, we get:
\begin{equation}
    \begin{array}{cl}
    P_e\left(d,w\right) \le
    0.5\sum\limits_{\tilde{w}=0}^{w}erfc\left(\sqrt{{r d
\gamma_{b}}+\frac{\tilde{L}(x)(w-2\tilde{w})}{2}}\right) {w \choose
\tilde{w}} \rho^{w-\tilde{w}} \times (1-\rho)^{\tilde{w}}
    \end{array}
 \label{eq6.51}
\end{equation}
%Consider now the following function:
%\begin{equation}
% F(x,y) = erfc(\sqrt{x+y}) - erfc(\sqrt{x})e^{-y}
%  \label{eq1lp1}
%\end{equation}
%with $x \ge 0$ and $y \ge 0$. By the definition of the $erfc$
%function it is easy to get:
%\begin{equation}
% \frac{\partial F(x,y)}{\partial x} = \frac{1}{2 \pi} e^{-x}e^{-y} \left(\frac{1}{\sqrt{x+y}}-\frac{1}{\sqrt{x}}\right)
%  \label{eq1lp2}
%\end{equation}
%Since $\frac{\partial F(x,y)}{\partial x} \le 0$ and $F(x,0) = 0$,
%we can obtain:
%\begin{equation}
% erfc(\sqrt{x+y}) \le erfc(\sqrt{x})e^{-y}
%  \label{eq1lp3}
%\end{equation}
To get the desirable simplification, consider now the Chernoff-Rubin
bound for the $erfc$ function, i.e.:
\begin{equation}
\begin{array}{c}
erfc(x) \le 2e^{-x^2}
\end{array}
 \label{eqCRB}
\end{equation}
Accordingly, we can write:
%\begin{equation}
%\begin{array}{c}
%P_e\left(d,w\right) \le 0.5 \sum\limits_{\tilde{w} = 0}^{w-1}
%erfc\left(\sqrt{{r d \gamma_{b}} + \frac{\tilde{L}(x)w}{2}
%-\tilde{L}(x)\tilde{w}}\right) \left(
%                                     \begin{array}{c}
%                                       w \\
%                                       \tilde{w} \\
%                                     \end{array}
%                                   \right)
%\rho^{w-\tilde{w}} \times (1-\rho)^{\tilde{w}}
%\end{array}
% \label{eq1AA2}
%\end{equation}
%
%
%it is now possible to get:
\begin{equation}
\begin{array}{c}
P_e\left(d,w\right) \le e^{-r d \gamma_b}
e^{-\frac{\tilde{L}(x)w}{2}} \sum\limits_{\tilde{w} = 0}^{w}
e^{\tilde{L}(x)\tilde{w}} \times {w \choose \tilde{w}}
\rho^{w-\tilde{w}} \times (1-\rho)^{\tilde{w}}
\end{array}
 \label{eq1AA3}
\end{equation}
which yields:
\begin{equation}
\begin{array}{c}
P_e\left(d,w\right) \le e^{-r d \gamma_b}
e^{-\frac{\tilde{L}(x)w}{2}}
\left[(1-\rho)e^{\tilde{L}(x)}+\rho\right]^w
\end{array}
 \label{eq1AA4}
\end{equation}
%, i.e., the upper bound in (\ref{eq1AA4}) is expected to be a good
%approximation for the pairwise error probability for $\gamma_b
%>> 1$.
Since $e^{\tilde{L}(x)/2} =
\sqrt{\frac{\tilde{\rho}}{1-\tilde{\rho}}}$, if we introduce the
term:
\begin{equation}
\begin{array}{c}
A = (1-\rho)\sqrt{\frac{\tilde{\rho}}{1-\tilde{\rho}}} + \rho
\sqrt{\frac{1-\tilde{\rho}}{\tilde{\rho}}}
\end{array}
 \label{eq1AA4.1.10}
\end{equation}
it is straightforward to get from (\ref{eq1AA4}):
\begin{equation}
\begin{array}{c}
P_e\left(d,w\right) \le e^{-r d \gamma_b} A^{w}
\end{array}
 \label{eq1AA4.1}
\end{equation}
The above expressions allows to separate the influence of signal to
noise ratio and codewords distance $d$ (first part) from the effect
of API (second part).
%In particular, if we assume
%perfect estimation, i.e., $\tilde{\rho} = \rho$, we get $A =
%2\sqrt{{{\rho}}({1-{\rho}})}$. In this case, the beneficial effect
%of a-priori information is reminiscent of the Chernoff bound for the
%evaluation of pairwise error probability in hard decoding. As a
%matter of fact, let us consider an equivalent BSC channel with
%transition probability (i.e., error probability) $1-\rho$. The
%a-priori probability ${\tilde{\pmb{x}}}$ can be seen as the output
%of the BSC channel to the actual transmitted sequence
%${\pmb{x}^{(t)}}$. The term $A^w$ in (\ref{eq1AA4.1}) is the
%Chernoff bound for the evaluation of the probability that
%${\pmb{x}^{(e)}}$ is closer than ${\pmb{x}^{(t)}}$ to the "observed"
%word ${\tilde{\pmb{x}}}$. Hence, the rationale for (\ref{eq1AA4.1})
%can be given as follows. The decoder makes a wrong decision when two
%conditions are jointly verified: a) the received signal belongs to
%the decision region of ${\pmb{c}^{(e)}}$; b) the a-priori
%information belongs to the decision region of
%${\pmb{x}^{(e)}}$.\\
A more precise measure of the pairwise error probability can be
derived by considering the exact evaluation of the first term in
(\ref{eq1AA4.1}) instead of its exponential bound, i.e.:
\begin{equation}
\begin{array}{c}
P_{e}\left(d,w\right) \simeq 0.5 erfc\left(\sqrt{r d
\gamma_b}\right) A^{w}
\end{array}
 \label{eq1AA4.12}
\end{equation}
Note that (\ref{eq1AA4.12}) gives an exact calculation of the
pairwise error probability for $\rho = \tilde{\rho} = 0.5$, i.e., in
absence of API. Even if (\ref{eq1AA4.12}) is not a strict bound for
$P_{e}\left(d,w\right)$, we will prove by simulations that it gives
a quite close upper bound
in most of the situations. \\
Equations (\ref{eq1AA4.1}) and (\ref{eq1AA4.12}) give rise to
interesting considerations about the properties of good channel
codes in presence of API. As in traditional codes' design, a good
code must be characterized by a high minimum Hamming weight $d$.
Moreover, in order to fully exploit the benefits of API, the code
structure should allow to associate information sequences with high
Hamming weights $w$ to codewords with low Hamming weights $d$. This
result can be easily understood if we
rewrite (\ref{eq1AA4.1}) as:\\
\begin{equation}
\begin{array}{c}
P_{e}\left(d,w\right) \le e^{-{r d \gamma_b}} e^{-{r \gamma_b}
ln\left({\frac{1}{A}}\right) \times w \frac{1}{r \gamma_b}} =
\left(e^{-{r \gamma_b}}\right)^{d+ w \times
ln\left({\frac{1}{A}}\right)\frac{1}{r \gamma_b}}
\end{array}
 \label{eq1AA5n}
\end{equation}
and if we observe that for reasonable $\rho$ estimates, i.e., $\rho
\cong \tilde{\rho}$, we get $A < 1$. Hence, denoting by $\alpha =
ln\left({\frac{1}{A}}\right)\frac{1}{r \gamma_b}$, a rule of the
thumb for designing good codes is that of maximizing the minimum
$d+w\alpha$ (with $\alpha > 0$). Of course, a rigorous analysis
should consider the trade-off between diminishing the pairwise error
probability from one side and increasing the number of bits in
errors $w$ from the other side.

\section{Uncoded Communications}

\begin{figure}
\begin{center}
\includegraphics[width=1\textwidth]{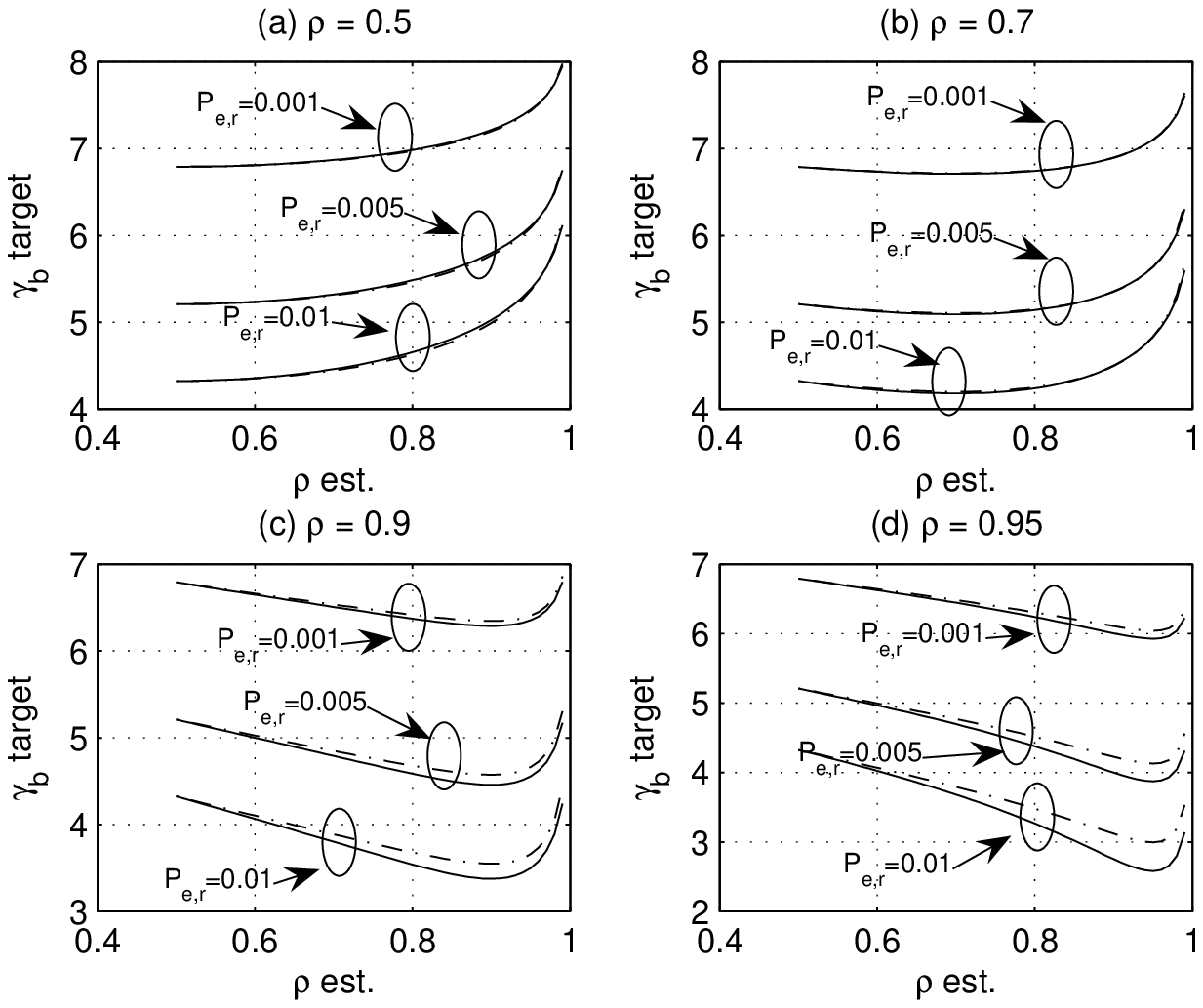}
\end{center}
\caption{$\gamma_b$ required to achieve $P_{e,r}$ versus the
estimated $\rho$ (i.e., ${\tilde{\rho}}$ is in the abscissa) in the
uncoded case: comparisons between exact calculation (straight lines)
and $P_{e,b}$ (dashed lines), for: (a) $\rho = 0.5$, (b) $\rho =
0.7$, (c) $\rho = 0.9$, (a) $\rho = 0.95$.} \label{Fig1 }
\end{figure}

In the uncoded case $r = k = n = 1$, $d = w = 1$ and the pairwise
error probability is equivalent to the bit error probability, which
can be derived according to (\ref{eq6.5}) as:
\begin{equation}
    \begin{array}{cl}
    P_e =
    0.5 erfc\left(\sqrt{{\gamma_{b}}\left(1+\frac{\tilde{L}(x)}{4
\gamma_{b}}\right)^2}\right) \rho + 0.5
erfc\left(\sqrt{{\gamma_{b}}\left(1-\frac{\tilde{L}(x)}{4
\gamma_{b}}\right)^2}\right) \times (1-\rho)
    \end{array}
 \label{eq6.5n}
\end{equation}
%Of course (\ref{eq6.5n}) is the exact bit error probability.
The
approximation (\ref{eq1AA4.12}) can be written in this case as:
\begin{equation}
\begin{array}{c}
P_{e,b}\left(d,w\right) \simeq 0.5 erfc\left(\sqrt{ \gamma_b}\right)
A
\end{array}
 \label{eq1AA4.121}
\end{equation}
A comparison between the exact calculation in (\ref{eq6.5n}) and the
approximation in (\ref{eq1AA4.121}) is given in Fig. \ref{Fig1 }. In
the y-axis we report the $\gamma_b$ required to achieve a target bit
error probability, say it $P_{e,r}$. In the x-axis we report
$\tilde{\rho}$. Four different $\rho$ values have been considered,
namely $\rho = 0.5$ in Fig. \ref{Fig1 } (a), $\rho = 0.7$ in Fig.
\ref{Fig1 } (b), $\rho = 0.9$ in Fig. \ref{Fig1 } (c) and $\rho =
0.95$ in Fig. \ref{Fig1 } (d). We note that approximation
(\ref{eq1AA4.121}) is almost exact for $\rho < 0.7$. Moreover, it is
a very close upper bound for $\rho > 0.7$ and for $P_{e,r} = 0.001$,
i.e., for high signal to noise ratios. As expected,
(\ref{eq1AA4.121}) gives a worse approximation for $\rho = 0.9$,
$\rho = 0.95$ and for $\gamma_b < 4$, where the bound (\ref{eqAA1})
is less tight. However, also in these cases, (\ref{eq1AA4.121})
gives a quite close upper bound for the bit error probability
evaluation. Hence, the proposed approximation allows to give a very
good measure of the performance gain which can be obtained by
exploiting API at the receiver, even in presence of imperfect
estimation. Note that the system performance is quite robust to
imperfect reliability estimation, at least for $\rho \le 0.95$. As
an example, for $\rho = 0.9$ and $P_{e,r} = 0.001$, an estimation of
$\tilde{\rho} = 0.8$ reduces the performance by only 0.1 dB with
respect to perfect estimation ($\tilde{\rho} = 0.9$), while an
estimation of $\tilde{\rho} = 0.95$ reduces the performance by less
than 0.08 dB. To sum up, results in Fig. \ref{Fig1 } show that API
allows to achieve reasonable performance gains at low $\gamma_b$
 with respect to the $\tilde{\rho} = 0.5$
case. This is true even in presence of not very accurate estimation
of the side information reliability $\rho$.

%As an example let us consider $\rho = 0.9$. In this case, the
%performance gain estimated by (\ref{eq1AA4.121}) in case of
%$\tilde{\rho} = \rho$ is of 0.46 dB for $P_{e,r} = 0.001$ (the
%approximation (\ref{eq1AA4.121}) estimates a gain of 0.45 dB)

%\begin{figure}
%\begin{center}
%\includegraphics[width=1\textwidth]{error_est2.eps}
%\end{center}
%\caption{Bit error probability versus the estimated $\rho$ (i.e.,
%${\tilde{\rho}}$ is in the abscissa) in the uncoded case:
%comparisons between exact calculation and upper bound, for $\rho =
%0.75$.} \label{Fig2 }
%\end{figure}
%\vfill
%\begin{figure}
%\begin{center}
%\includegraphics[width=1\textwidth]{error_est3.eps}
%\end{center}
%\caption{Bit error probability versus the estimated $\rho$ (i.e.,
%${\tilde{\rho}}$ is in the abscissa) in the uncoded case:
%comparisons between exact calculation and upper bound, for $\rho =
%0.9$.} \label{Fig3 }
%\end{figure}
%\vfill

\section{Coded Communication Schemes}

\subsection{Convolutional codes}

\begin{figure}
\begin{center}
\includegraphics[width=1\textwidth]{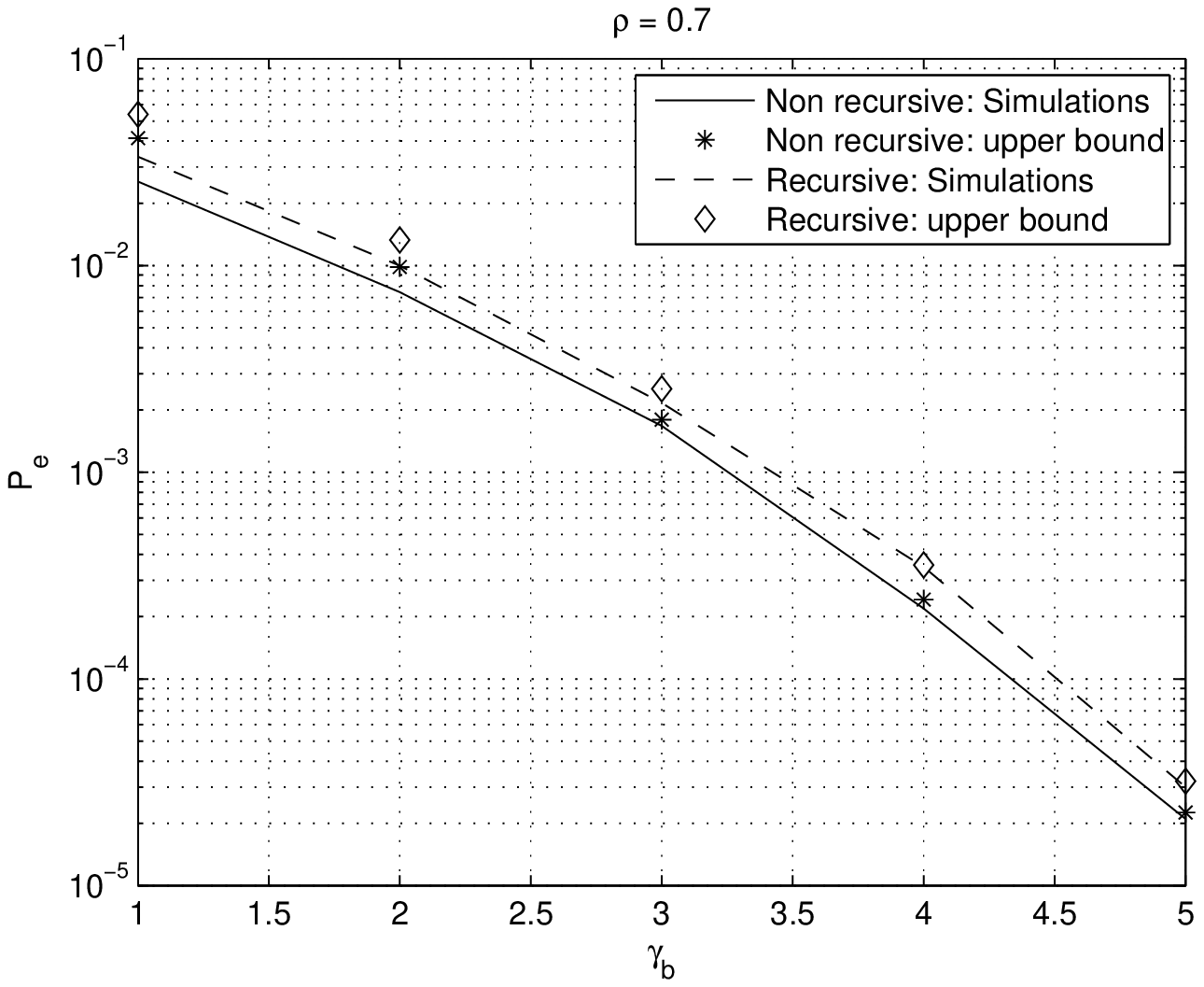}
\end{center}
\caption{$P_e$ versus $\gamma_b$ comparisons between recursive and
non recursive convolutional codes: simulation results are shown
together with the ${P}_{e,1}$ upper bounds for $\rho =
0.7$}\label{Fig2}
\end{figure}

\begin{figure}
\begin{center}
\includegraphics[width=1\textwidth]{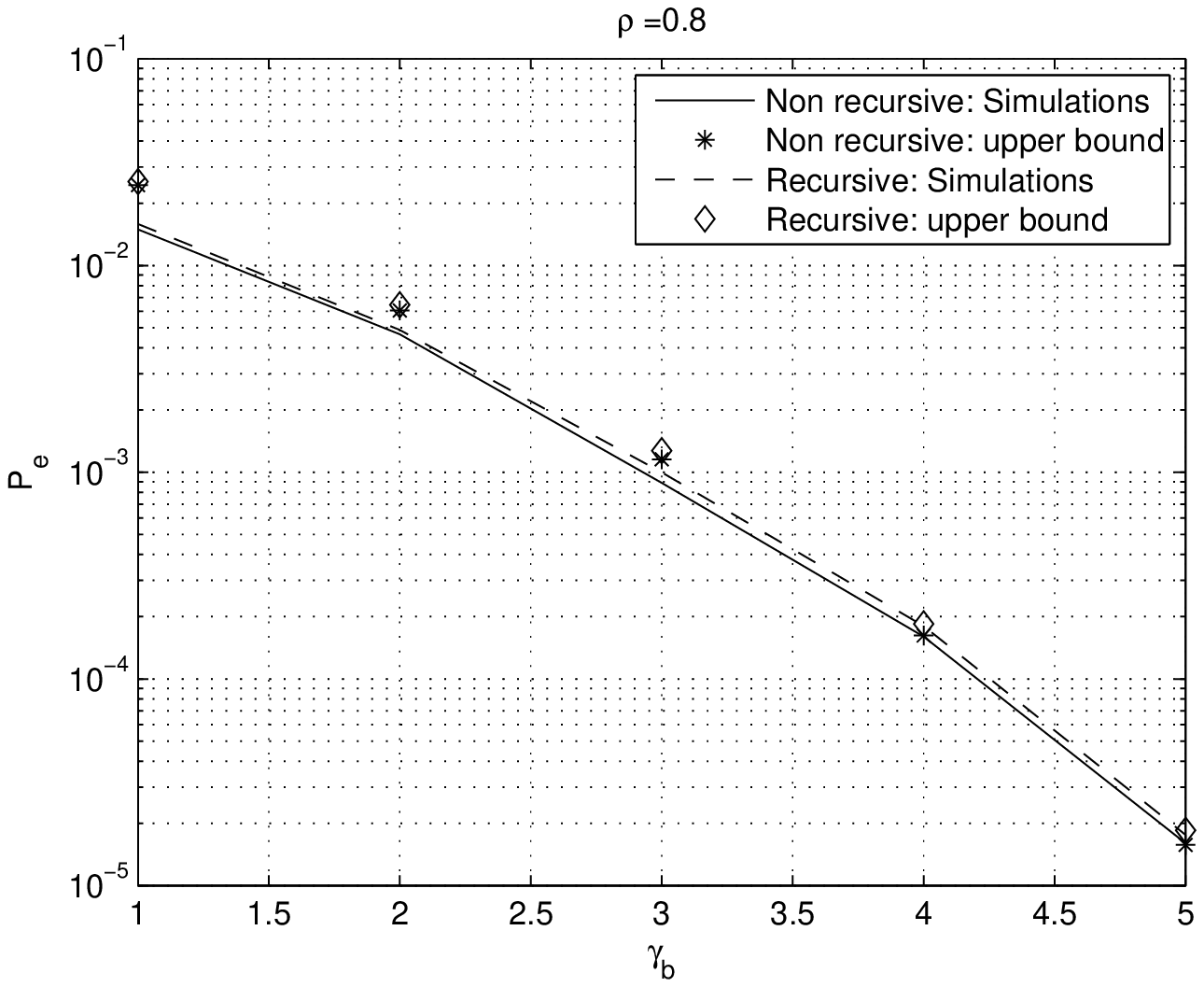}
\end{center}
\caption{$P_e$ versus $\gamma_b$ comparisons between recursive and
non recursive convolutional codes: simulation results are shown
together with the ${P}_{e,1}$ upper bounds for $\rho =
0.8$}\label{Fig3}
\end{figure}

\begin{figure}
\begin{center}
\includegraphics[width=1\textwidth]{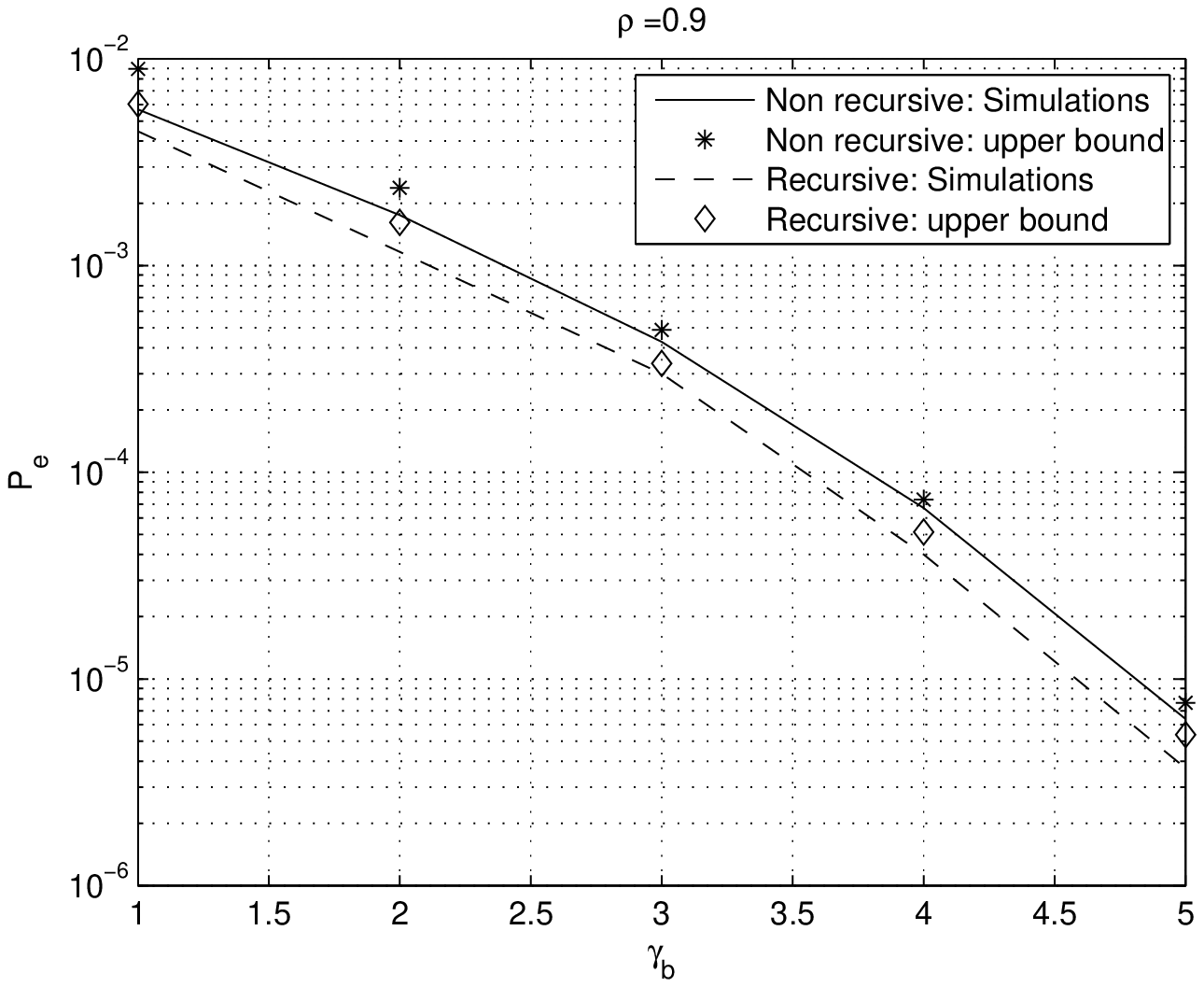}
\end{center}
\caption{$P_e$ versus $\gamma_b$ comparisons between recursive and
non recursive convolutional codes: simulation results are shown
together with the ${P}_{e,1}$ upper bounds for $\rho =
0.9$}\label{Fig4}
\end{figure}

\begin{figure}
\begin{center}
\includegraphics[width=1\textwidth]{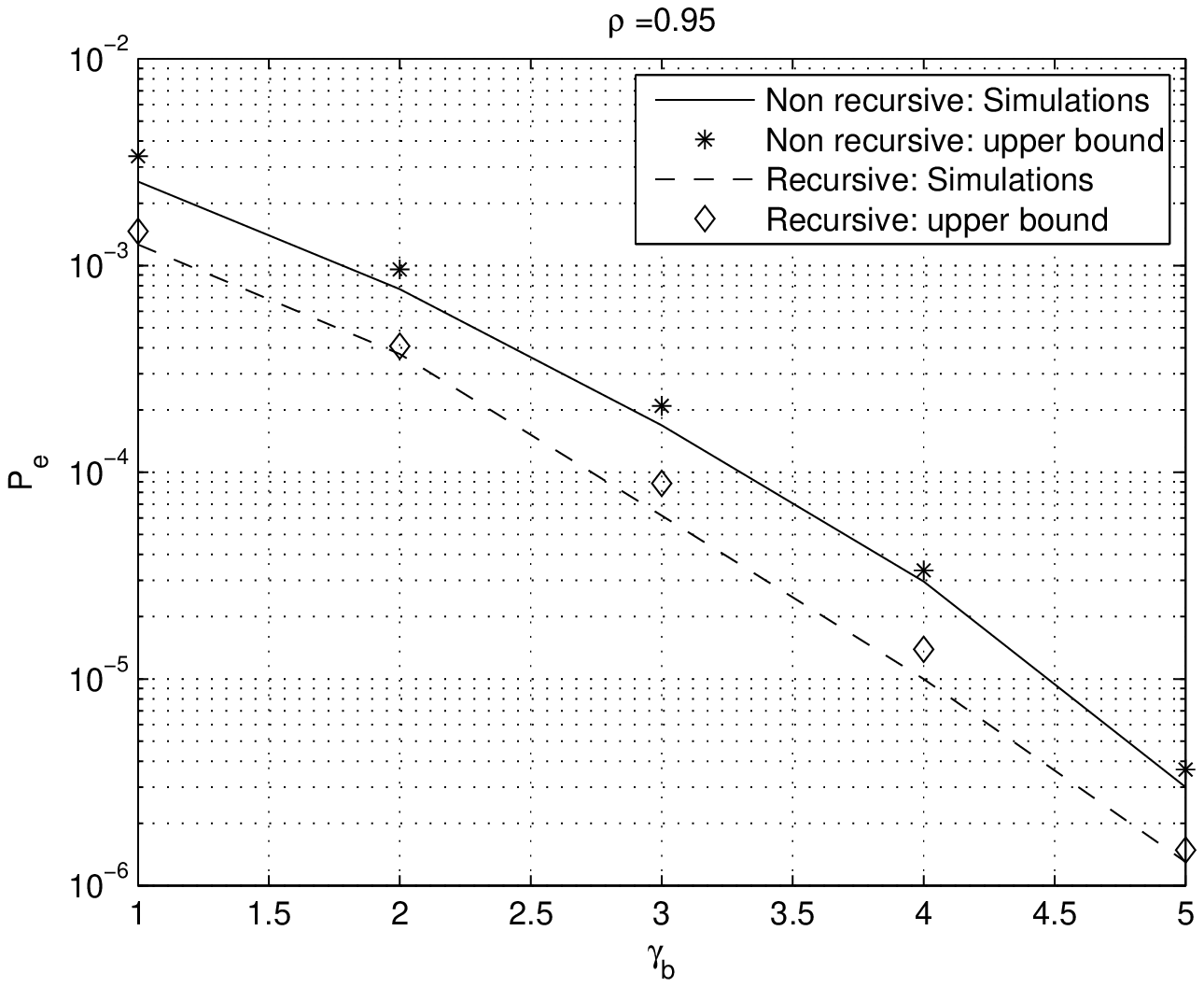}
\end{center}
\caption{$P_e$ versus $\gamma_b$ comparisons between recursive and
non recursive convolutional codes: simulation results are shown
together with the ${P}_{e,1}$ upper bounds for $\rho =
0.95$}\label{Fig5}
\end{figure}

\begin{figure}
\begin{center}
\includegraphics[width=1\textwidth]{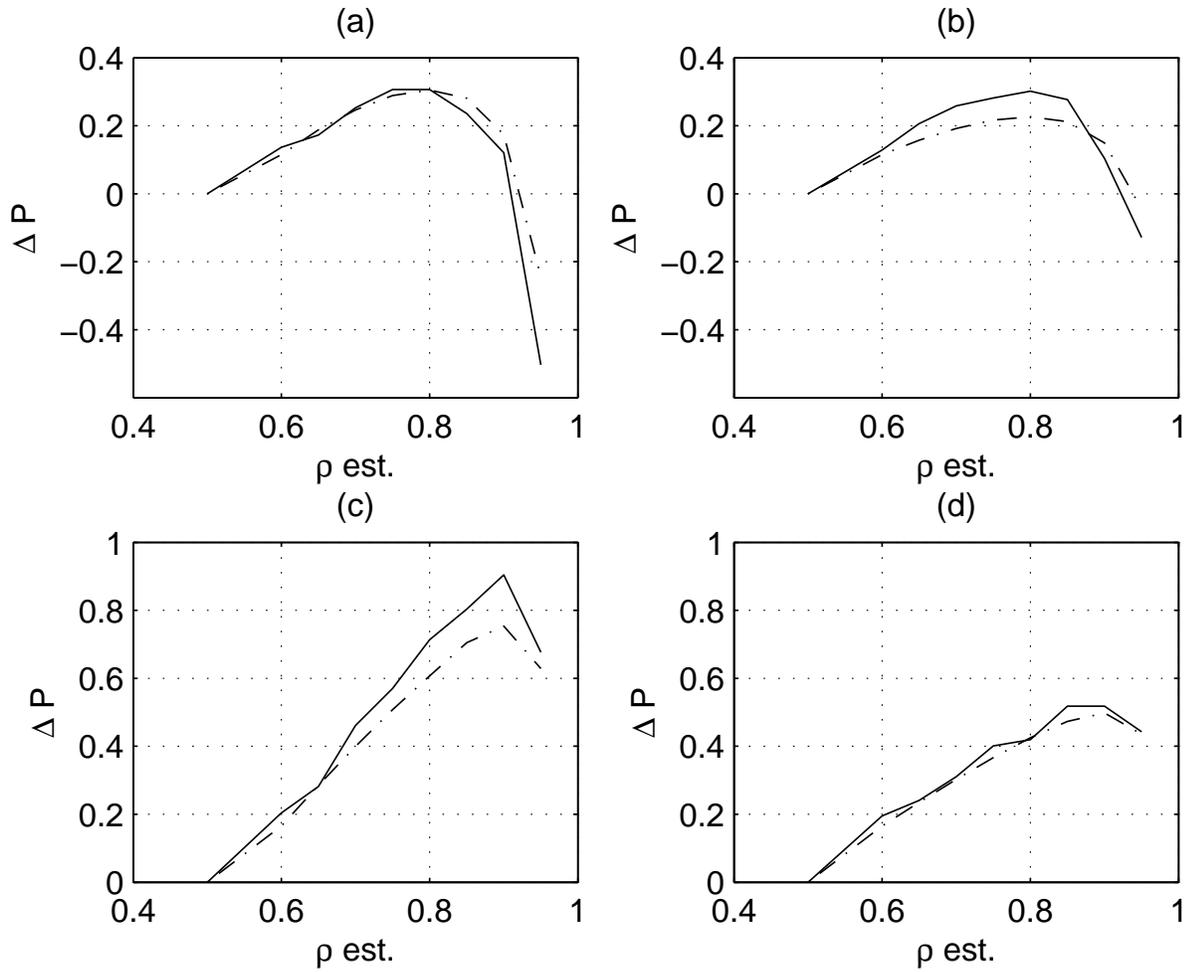}
\end{center}
\caption{$\Delta P$ versus the estimated $\rho$ (i.e.,
${\tilde{\rho}}$ is in the abscissa) for ${P}_{e,r} = 0.0001$:
comparisons between simulations (straight lines) and analysis in
(\ref{eq13AA25.new}) (dashed lines), for: (a) $\rho = 0.8$,
recursive code (b) $\rho = 0.8$, non recursive code (c) $\rho =
0.9$, recursive code (d) $\rho = 0.9$, non recursive code
.}\label{Fig6}
\end{figure}

Convolutional coding schemes \cite{Sklar}, \cite{Proakis} allow an
easy coding implementation with very low power and memory
requirements and, hence, they seem to be particularly suitable for
utilization in WSNs \cite{Karl}. Moreover, as stated in the
Introduction, correlation among sources may be directly converted to
API at the receiver. Hence, optimum decoding schemes can be easily
derived by including the a-priori probabilities in the branch
metrics of the Viterbi algorithm
according to equation (\ref{eq4AA}).\\
As in traditional convolutional coding (i.e., without API), it is
possible to derive an upper bound of the bit error probability as
the weighted \footnote{The weights are the information error
weights} sum of the pairwise error probabilities relative to all
paths which diverge from the zero state and marge again after a
certain number of transitions \cite{Sklar}. This is possible because
of the linearity of the code and because the pairwise error
probability (\ref{eq6.5}) depends only on the weights $d$ and $w$,
and not on the actual
transmitted sequence.\\
In particular, it is possible to evaluate the input-output transfer
function $T(W,D)$ by means of the state transition relations over
the modified state diagram \cite{Sklar}. The generic form of
$T(W,D)$ is:
\begin{equation}
    \begin{array}{cl}
T(W,D) = \sum\limits_{w,d}^{} \beta_{w,d} W^w D^d
    \end{array}
 \label{eq1BB}
\end{equation}
where $\beta_{w,d}$ denotes the number of paths that start from the
zero state and reemerge with the zero state and that are associated
with an input sequence of weight $w$, and an output sequence of
weight $d$. Accordingly, we can get an upper bound of the bit error
probability as:
\begin{equation}
    \begin{array}{cl}
    {P}_{e,b} \le \sum\limits_{w,d} \beta_{w,d} \times w \times {P}_e\left(d,w\right)
    \end{array}
 \label{eq13AA1}
\end{equation}
where $P_e(d,w)$ is the pairwise error probability. Let now denote
by $P_{e,e}$ the exact pairwise error probability derived in
(\ref{eq6.5}) and by $P_{e,a}$ the approximation (\ref{eq1AA4.12}).
Accordingly, we get the following bound for the bit error
probability:
\begin{equation}
    \begin{array}{cl}
    {P}_{e,1} = \sum\limits_{w,d} \beta_{w,d} \times w \times {P}_{e,e}\left(d,w\right)
    \end{array}
 \label{eq13AA21}
\end{equation}
%and
%\begin{equation}
%    \begin{array}{cl}
%    {P}_{e,2} = \sum\limits_{w,d} \beta_{w,d} \times w \times {P}_{e,a}\left(d,w\right)
%    \end{array}
% \label{eq13AA22}
%\end{equation}
A second bound can be obtained by considering the loose upper bound
(\ref{eq1AA4.1}):
\begin{equation}
    \begin{array}{cl}
    {P}_{e,2} = \sum\limits_{w,d} \beta_{w,d} \times w \times e^{-r d \gamma_b}
A^{w}
    \end{array}
 \label{eq13AA23}
\end{equation}
From (\ref{eq1BB}) and (\ref{eq13AA23}) it is straightforward to
obtain:
\begin{equation}
    \begin{array}{cl}
    {P}_{e,2} = A \times \frac{\partial T(W,D)}{\partial W}|_{W = A, D = e^{-r \gamma_b}}
    \end{array}
 \label{eq13AA25}
\end{equation}
Since ${P}_{e,2}$ is a monotone decreasing function of $\gamma_b$,
it is straightforward to carry out numerical inversion of
$(\ref{eq13AA25})$ with respect to $\gamma_b$. Such an inversion
allows to get an estimation of the threshold signal-to-noise ratio
$\gamma_b(P_{e,r},A)$ corresponding to a given ${P}_{e,2} =
P_{e,r}$. Note that $\gamma_b(P_{e,r},1)$ corresponds to the
threshold $\gamma_b$ when no API is present at the receiver.
Accordingly, the signal-to-noise-ratio gain due to API can be
derived as:
\begin{equation}
    \begin{array}{cl}
    \Delta P =  \frac{\gamma_b(P_{e,r},1)}{\gamma_b(P_{e,r},A)}
    \end{array}
 \label{eq13AA25.new}
\end{equation}
In order to assess the validity of the previous analysis, we have
carried out computer simulations for both recursive and
non-recursive convolutional codes. In both cases, we have considered
a rate $r = 0.5$ and a constraint length $K = 4$. Hence, the codes
can be univocally characterized by the generator polynomials
$G^{(1)}(D) = g^{(1)}_{3} D^{3}+g^{(1)}_{2}
D^{2}+g^{(1)}_{1}D+g^{(1)}_{0}$, $G^{(2)}(D) = g^{(2)}_{3} \times
D^{3}+g^{(2)}_{2} D^{2}+g^{(2)}_{1}D+g^{(2)}_{0}$ and by the
feedback polynomial $H(D) = h_{3} \times D^{3}+h_{2}
D^{2}+h_{1}D+h_{0}$. As for the non-recursive code we have
considered the maximum $d_{free}$ code which is optimum in the
uncorrelated scenario \cite{Proakis}, i.e., $G^{(1)}(D) =
D^3+D^2+1$, $G^{(2)}(D) = D^3+D^2+D+1$ and, of course, $H(D) = 1$.
Such a code is characterized by a transfer function:
\begin{equation}
    \begin{array}{cl}
    T(D,W) =  \frac{D^6 W^2 + D^7 W -D^8 W^2}{1-2DW-D^3W} = D^6 W^2
    + D^7 W + 2 D^7 W^3 + \ldots
    \end{array}
 \label{eqTF1}
\end{equation}
It is worth noting that the non-recursive code is characterized by a
path with minimum distance $d_{free} = 6$ and information weight $w
= 2$. \\
As for the recursive code, we consider the generator polynomials
$G^{(1)}(D) = D^3+D+1$, $G^{(2)}(D) = D^3+D^2+D+1$ and $H(D) =
D^3+D^2+1$. Such a code is characterized by a transfer function:
\begin{equation}
    \begin{array}{cl}
    T(D,W) = \frac{D^6 W^2 (D^6 W^4 - 2D^6 W^2 +D^6 +D^5 W -D^5 W^3 + 2W^4 D^2 - 2D^2 W^2 -D^2 W^6 +D W^5 -2DW^3 +2DW +W^2)}
    {1-D^8 W^4 + 2D^8 W^2 -D^8 -D^7 W +D^7 W^3 -D^5 W +D^5 W^3 +D^4 W^2 -2 D^4 W^4 +D^4 W^6 -D^3 W^5 +2 W^3 D^3 -2 D^3 W -2 D W}
    = \\ = D^6 W^4
    + D^7 W^7
     + 2 D^7 W^3 + \ldots
    \end{array}
 \label{eqTF1}
\end{equation}
The recursive code is characterized by a path with minimum distance
$d_{free} = 6$ and information weight $w = 4$. \\
Given the above, for high signal to noise ratios the bit error
probability can be approximated as $P_e \cong A^2 erfc\left(\sqrt{3
\gamma_b}\right)$ for the non-recursive code and as $P_e \cong 2 A^4
erfc\left(\sqrt{3 \gamma_b}\right)$ for the recursive code.
Accordingly, we expect that the recursive code outperforms the
non-recursive one for $A < \frac{1}{\sqrt{2}}$. Under the hypothesis
of perfect reliability estimation, i.e., $\rho = \tilde{\rho}$, this
means that the recursive code performs better for $\rho
> 0.85$.\\
Comparisons between the above codes are shown in Figs.
\ref{Fig2}-\ref{Fig5} in the case of perfect reliability estimation
and for different $\rho$ values, namely $\rho = 0.7$ in Fig.
\ref{Fig2}, $\rho = 0.8$ in Fig. \ref{Fig3}, $\rho = 0.9$ in Fig.
\ref{Fig4} and $\rho = 0.95$ in Fig. \ref{Fig5}. In all figures
simulation results are shown together with the ${P}_{e,1}$ upper
bound derived in
(\ref{eq13AA21}).\\
As one can observe, the analytical upper bound derived in
${P}_{e,1}$ is quite tight and, in particular, tends to perfectly
match simulation results for high signal to noise ratios. Moreover,
as expected, the recursive code clearly outperforms the
non-recursive one for $\rho \ge 0.9$, while for $\rho < 0.8$ the non
recursive code performs better.\\
More extensive comparisons between simulations and theoretical
analysis have been carried out to evaluate the signal to noise ratio
gain $\Delta P$ which can be obtained by means of API at the
receiver. Such results are shown in Fig. \ref{Fig6}, where $\Delta
P$ versus ${\tilde{\rho}}$ for ${P}_{e,r} = 0.0001$ is shown.
Simulation results are the straight lines while analytical results
derived according to (\ref{eq13AA25.new}) are the dashed lines.
Different $\rho$ values have been considered, namely $\rho = 0.8$ in
Figs. \ref{Fig6} (a) and \ref{Fig6} (b) and $\rho = 0.9$ in Figs.
\ref{Fig6} (c) and \ref{Fig6} (d). Eventually, results for the
recursive code are shown in Figs. \ref{Fig6} (a) and \ref{Fig6} (c)
and results for the non recursive code are shown in Figs. \ref{Fig6}
(b) and \ref{Fig6} (d). We note that approximation (\ref{eq13AA25})
allows to predict quite well the beneficial effect of a priori
information even in the case of non perfect $\rho$ estimation. It is
worth noting that, as expected, recursive code takes grater
advantage from exploiting a priori information than non recursive
code. As an example, for $\rho = 0.9$ the maximum performance gain
(i.e., the performance gain which is obtained for $\rho =
{\tilde{\rho}}$) is 0.9 dB for the recursive code and 0.5 dB for the
non recursive code. On the other hand, the recursive code is much
more sensitive to estimation errors than the non recursive code (on
account of the higher minimum $w$).

\subsection{Random Selection Of Codes}

In an attempt to derive a general framework for the evaluation of
the impact of API in the performance of coded signals, we now
consider random selection of codes and we evaluate a bound on the
average bit error probability. In the proposed approach we extend
the considerations made in \cite{Proakis}, Section 7-2, to the case
of a priori information at the receiver. In particular, denoting by
$M = 2^k$, we consider the ensemble of $(2^n)^M$ distinct ways in
which we can select $M$ binary codewords from the available $2^n$
words of length $n$. Each code selection leads to a different
communication system which is characterized by its probability of
error. As done in \cite{Proakis} we assume that the choice of $M$
codewords is based on random selection. In particular, in
\cite{Proakis} it is derived an upper bound on the expected pairwise
error probability for a given Hamming distance $d$ as:
\begin{equation}
\begin{array}{c}
\overline{P_e} \le \frac{1}{2^n}\sum\limits_{d = 0}^{n}
                                     {n \choose d}
                                   e^{-r d \gamma_b}\\

\end{array}
 \label{eq1AA3.1}
\end{equation}
where the average is evaluated over the ensemble of $(2^n)^M$ codes.
Let now consider the upper bound derived in (\ref{eq1AA4.1}) for the
pairwise error probability in presence of API. It is worth noting
that in this case the pairwise error probability depends on $d$ and
$w$, whereas in absence of API it depends only on $d$. Moreover,
since the code selection is random, $d$ and $w$ are binomial
independent random discrete variables. Hence, averaging over the
ensemble of $(2^n)^M$ codes we get in this case:
\begin{equation}
\begin{array}{c}
\overline{P_e} \le \frac{1}{2^n}\frac{1}{2^k}\sum\limits_{d =
0}^{n}\sum\limits_{w = 0}^{k}{n \choose d} {k \choose w}
                                   \left(e^{-r \gamma_b}\right)^d A^w
\end{array}
 \label{eq1AA3.2}
\end{equation}
where $A$ is defined in (\ref{eq1AA4.1.10}). From the above, it is
then straightforward to derive:
\begin{equation}
\begin{array}{c}
\overline{P_e} \le \left(\frac{1+e^{-r \gamma_b}}{2}\right)^n \times
\left(\frac{1+A}{2}\right)^k
\end{array}
 \label{eq1AA3.3}
\end{equation}
Eventually, since the average pairwise error probability is
independent of $d$ and $w$ we can easily obtain an union bound on
the average bit error probability by considering the sum of all the
$M-1$ possible error events, i.e.:
\begin{equation}
\begin{array}{c}
\overline{P_{e,b}} \le (M-1)\left(\frac{1+e^{-r
\gamma_b}}{2}\right)^n \times \left(\frac{1+A}{2}\right)^k < M
\left(\frac{1+e^{-r \gamma_b}}{2}\right)^n \times
\left(\frac{1+A}{2}\right)^k
\end{array}
 \label{eq1AA3.4}
\end{equation}
This result can be expressed in a more convenient form by
introducing the terms $R_1 = {log_{2}\left(\frac{2}{1+e^{-r
\gamma_b}}\right)}$ and $\eta =
{log_{2}\left(\frac{2}{1+A}\right)}$. Accordingly, since $M = 2^k$
and $r = k/n$, (\ref{eq1AA3.4}) becomes:
\begin{equation}
\begin{array}{c}
\overline{P_{e,b}} < 2^{nr-nR_1-nr\eta} =
2^{-n\left[R_1-r(1-\eta)\right]}
\end{array}
 \label{eq1AA3.7}
\end{equation}
We have thus obtained a similar expression for the average bit error
probability as that in \cite{Proakis}, with the introduction of the
term $\eta$ which takes into account the effect of API. Hence,
introducing the cutoff rate $R_0 = \frac{R_1}{1-\eta}$ we conclude
that when $r < R_0$ the average bit error probability
$\overline{P_{e,b}} \rightarrow 0$ as the code length $n \rightarrow
\infty$, i.e., there exist "good" codes that have a probability of
error
which goes to zero. \\
In order to derive a measure of the performance gain which can be
obtained by API, we introduce the term $\gamma_{b,t}$ as the minimum
$\gamma_b$ which ensures the presence of a good codes for a given
transmission rate $r$. It is straightforward to derive from the
above:
\begin{equation}
\begin{array}{c}
\gamma_{b,t}(\eta) =
\frac{1}{r}ln\left(\frac{1}{2^{1-r(1-\eta)}-1}\right)
\end{array}
 \label{eq1AA3.9}
\end{equation}
The signal-to-noise-ratio gain due to API for a given $r$ can be
evaluated in this case as:
\begin{equation}
    \begin{array}{cl}
    \Delta P =  \frac{\gamma_{b,t}(0)}{\gamma_{b,t}(\eta)} =
    \frac{ln\left({2^{1-r}-1}\right)}{ln\left({2^{1-r(1-\eta)}-1}\right)}
    \end{array}
 \label{eq13AA25.1}
\end{equation}
It is now possible to get an insight into the performance of $r =
0.5$ convolutional codes presented in the previous Section where the
$\Delta P$ gain has been evaluated for a target $P_{e,r} = 0.0001$.
In particular, considering the case $\rho = \tilde{\rho} = 0.9$
(i.e., $\eta \cong 0.3219$) and setting $r=0.5$, we get from
(\ref{eq13AA25.1}) $\Delta P \cong 2.1$ dB, whereas the recursive
convolutional code proposed in the previous Section yields $\Delta P
\cong 0.9$ dB and the non recursive convolutional codes yields
$\Delta P \cong 0.5$ dB (see Fig. \ref{Fig6}).

\subsection{Turbo codes}

\begin{figure}
\begin{center}
\includegraphics[width=1\textwidth]{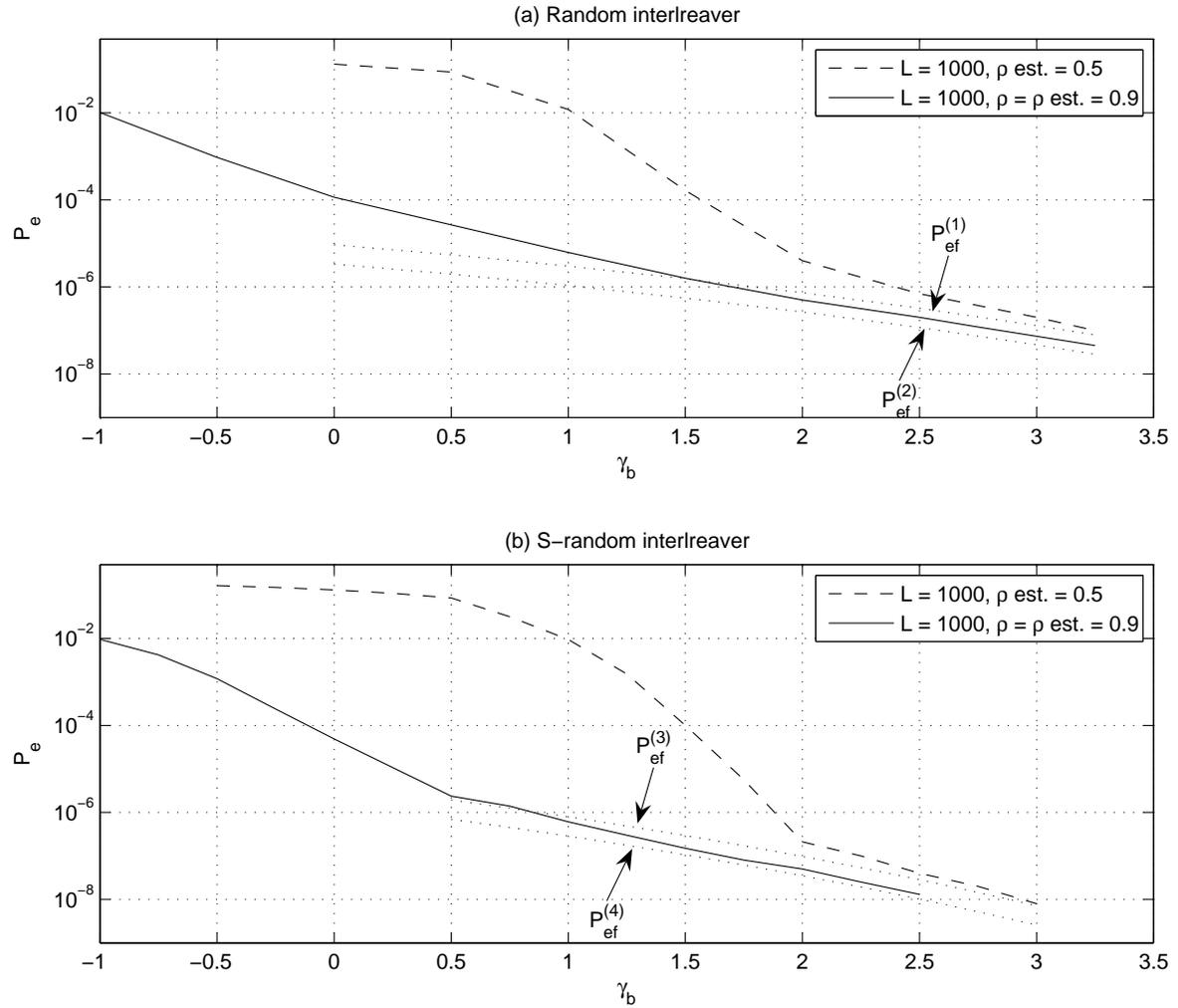}
\end{center}
\caption{$P_e$ versus $\gamma_b$ for $TC$ with $L = 1000$, random
interleaving (a) and S-random interleaving (b): comparisons between
no a-apriori ($\rho = 0.5$) and a-priori with $\rho = \rho$ est. $=
0.9$.}\label{Fig7}
\end{figure}

In an attempt of reducing the gap between the theoretical $\Delta P$
derived in the previous Section and the actual $\Delta P$ which can
be obtained by real codes, we analyze in this subsection the
performance of parallel concatenated codes (turbo codes
\cite{Turbo1}, \cite{Turbo2}) in presence of API at the decoder. As
it is well known, the trick in turbo coding is
%to
%match low-weight encodings of one permutation with high-weight
%encodings of the other permutation. This allows to produce total
%weights significantly higher than the low weights that are possible
%from each of the simple component codes individually. This goal can
%be achieved even with random permutations (random interleaver),
%provided that the frame size is sufficiently high (greater than
%1000) and the constituent codes are recursive convolutional codes.
%Indeed, in this case low weight codewords are "statistically" broken
to "statistically" break low weight codewords by means of random
interleaving, so that the performance of the decoder in the region
of \emph{not too much low} BERs \footnote{Not too much low BERs mean
before approaching the well known error floor region of turbo
codes.} is mainly driven by high weights codewords (which occur with
much higher probability than low weights codewords). On the other
hand, since constituent codes are convolutional codes, high weights
codewords are also characterized by high information weight (i.e.,
high $w$ values). Hence, in this BER region, we expect that turbo
codes allow to take the best advantage of exploiting API at the
receiver. On the contrary, for random interleaving, the performance
of Turbo codes at very low BERs is mainly dominated by low distance
codewords \cite{Turbo3}. Such codewords are also characterized by
small $w$ values and hence we expect that in the error floor region
the gain which can be obtained by
exploiting API is small, i.e., similar to the gain that can be obtained by convolutional codes. \\
To elaborate, let us consider a two-code turbo code with random
interleaving and with identical constituent convolutional encoders.
As it is discussed in \cite{Turbo1}, the weight 2 (i.e., $w = 2$)
input data sequences which correspond to low weight codewords are
the sequences which dominate the performance at low BER values.
%This
%is because higher weights input sequences have smaller probabilities
%of being reproduced to the other encoder with the same low weight
%codewords.
Let us denote by $d_2$ the minimum codewords' weight which
correspond to single error events of weight $w=2$ in the trellis of
the constituent codes. The minimum weight of the turbo code's
codewords which corresponds to such $w=2$ sequences is $d_{2,t} =
2d_2-2$. This distance is obtained when the same error event is
presented at the input of the two encoders (it is two times $d_2$
minus the information weight $w$, since the
 systematic bits are sent only once). The bit error
probability of two-codes turbo codes in the error floor region,
namely $P_{ef}^{(1)}$, can then be approximated as:
\begin{equation}
    \begin{array}{cl}
    P_{ef}^{(1)} \cong 2 K_1 0.5 erfc\left(\sqrt{r\gamma_b
    d_{2,t}}\right) = K_1 erfc\left(\sqrt{r\gamma_b d_{2,t}}\right)
 \label{eq13AA25.2}
 \end{array}
\end{equation}
where $K_1$ is the number of turbo coded sequences with information
weight $w = 2$ and codeword's weight $d_{2,t}$. For random
interleaving it can be easily shown that $K_1 = 2/k$ \cite{Turbo1}.
According to the analysis provided in the previous Sections, we then
expect that the bit error probability in presence of API, namely
$P_{ef}^{(2)}$, is $A^2$ smaller than $P_{ef}^{(1)}$, i.e.:
\begin{equation}
    \begin{array}{cl}
    P_{ef}^{(2)} \cong \frac{2}{k} erfc\left(\sqrt{r\gamma_b
    d_{2,t}}\right) \times A ^ 2
 \label{eq13AA25.3}
 \end{array}
\end{equation}
where $A$ is defined in (\ref{eq1AA4.1.10}). \\
As it is well known, performance of turbo codes can be improved by a
more accurate design of the interleaver \cite{Turbo3}. As an
example, S-random interleavers \cite{Turbo1} allow to avoid short
cycle events, i.e., two bits which are close to each other both
before and after interleaving. For comparison purposes, we then
consider a specific interleaver derived by applying the S-random
algorithm.\\
Computer simulations of a two-code turbo code system with both
random and S-random interleavers have then been carried out. The
constituent codes are $r=1/2$ recursive convolutional codes with
constraint length $K = 4$, $G^{(2)}(D) = D^{3}+ D^{2}+1$, $H(D) =
D^{3}+ D+1$, and $G^{(1)}(D) = H(D)$, (systematic code). The overall
rate of the turbo code is $r = 1/3$ which is increased to $r = 1/2$
via classical puncturing technique which enables to select the coded
bits alternatively from the two encoders. The algorithm used by the
two convolutional decoders at the receiver is based on the MAP BCJR
scheme \cite{BCJR}, which allows the inclusion of API in the form of
LLRs of
the input data.\\
Fig. \ref{Fig7} show the BER versus $\gamma_b$ for the turbo codes
($TC$) introduced above. The frame size $k$ of the information
sequence (i.e., the interleaving size) is set to $k = 1000$ bits and
the maximum number of iterations of turbo decoding is set to 10.
Performance of random (\ref{Fig7} (a)), and S-random (\ref{Fig7}
(b)) interleavers are shown for the case of no API, i.e., $\rho =
0.5$, and API with $\rho = 0.9$ and perfect estimation, i.e., $\rho
= \tilde{\rho}$. Theoretical curves for the random interleaving
evaluated according to (\ref{eq13AA25.2}) and (\ref{eq13AA25.3}) are
also shown. Note that for the considered code, $K_1 = 2/1000 =
0.002$.
%while $K_2$ is
%evaluated by fitting the simulation curves for $\rho = 0.5$ in the
%error floor region. Such a fitting gives $K_2 = 0.17$.
As far as
$d_{2,t}$ is concerned,
%note that in this case the systematic part of the codeword has
%weight 2. Hence, by denoting $d_2$ the minimum distance codeword of
%the constituent convolutional code which corresponds to a single
%error event of weight $w=2$, we get that the minimum weight of the
%parity bits of the unpunctured two-codes turbo codeword is $p_{2,1}
%= 2(d_2-2)$.
on account of puncturing we get $d_{2,t} = d_2$.
%In a similar way,
%it is straightforward to derive $d_{2,2,t} = 2d_2$.
The distance $d_2$ can be easily computed by means of the modified
state diagram \cite{Sklar}. In particular, for the considered
constituent codes we have $d_2 = 8$, which yields $d_{2,t} = 8$.
 %and
%$d_{2,2,t} = 16$.
%Eventually, since $\tilde{\rho} = \rho = 0.9$, we get $A = 0.6$.
Eventually, we also show the theoretical curves for the S-random
case. In this case a performance analysis in the error floor region
can be provided by following the WSE method proposed in
\cite{Turbo4}, where an union bound of the bit error probability is
calculated as the partial sum of the dominant terms (corresponding
to small code weights). Of course, we can also straightforwardly
derive the bit error probability in presence of API by multiplying
each term of the upper bound's partial sum by $A^w$, $w$ being the
information weight of this term. Theoretical curves
for the S-random case are denoted in Fig. \ref{Fig7} by $P_{ef}^{(3)}$, for the $\tilde{\rho} = 0.5$ case, and $P_{ef}^{(4)}$, for the $\tilde{\rho} = \rho = 0.9$ case.\\
Several comments can be drawn by the curves shown in Fig.
\ref{Fig7}. First of all note that, as expected, S-random
interleaver allows to achieve performance better than random
interleaver. Moreover, for $BER \ge 10^{-5}$ the considered turbo
codes allow to exploit API much better than convolutional codes
considered in the previous Section. As an example, if we consider
$P_{e,r} = 10^{-4}$ we observe that the performance gain due to API
is higher than $1.6$ dB for S-random interleaver and slightly lower
than $1.5$ dB for random interleaver \footnote{Remember that
recursive convolutional codes considered in this paper were able to
achieve a performance gain of 0.9 dB}. Similar gains are still
achieved for $P_{e,r} = 10^{-5}$. This result is due to the fact
that error events which mainly occur for such medium BER values are
characterized by high $w$ values. Instead, as expected, in the error
floor region the curves for $\rho = 0.5$ and $\rho = \tilde{\rho} =
0.9$ get closer since in this case the performance behavior is
determined by low $w$ error events. It is also worth noting that the
error floor fittings are very close to simulation results, thus
confirming the validity of the proposed analysis.\\
Results in Fig. \ref{Fig7} suggest that an accurate design of the
interleaver in turbo codes may help the decoder to exploit better
the API (if there is any). In particular, since the constituent
codes of turbo codes are convolutional codes, the possibility of
avoiding small $w$ codewords is fully demanded to the possibility of
the interleaver to break small weight input data sequences. Hence,
even if the design of optimal interleavers in presence of API is out
of the scope of this work, we can conclude that good interleaver for
the classical case (no API) are good also for the case of API at the receiver.\\
A question which arises from previous comments is wether turbo codes
allow to approach the performance gain $\Delta P$ which has been
derived in the previous Section for infinite length random codes. Of
course the performance gain depends in general on the target BER
$P_{e,r}$ that can be accepted. If we consider $P_{e,r} = 10^{-5}$
we see from Fig. \ref{Fig7} that such a BER is quite close to the
error floor region. To increase the $\Delta P$ for such a BER is
then necessary to lower the error floor region, i.e., to decrease
the probability of the occurrence of low $w$ error events. As it is
well known from the literature \cite{Turbo2} this can be easily
obtained by increasing the frame size $k$. Hence we have run
computer simulations for different $k$ and for the S-random
interleaver. Results are summarized in Fig. \ref{Fig8} where $\Delta
P$ versus ${\tilde{\rho}}$ for ${P}_{e,r} = 10^{-5}$ is shown for
$\rho = 0.7$ (Fig. \ref{Fig8} (a)), $\rho = 0.9$ (Fig. \ref{Fig8}
(b)) and for different $k$ values, namely $k = 100$, $k = 1000$, and
$k = 100000$. For comparison purposes, we also show $\Delta P$ of
random codes ($RC$) with $k = \infty$  obtained through equation
(\ref{eq13AA25.1}). Note that as $k$ increases up to $100000$, the
performance gain due to API of $TC$s approach the theoretical gain
of infinite length $RC$s. Of course this is true for ${P}_{e,r} =
10^{-5}$ while, for the considerations drawn before, it could not be
true anymore for a lower BER target. It is also worth noting that
the theoretical analysis for $RC$s gives an accurate bound of the
allowable gains that can be obtained by exploiting API at the
receiver even in presence of estimation errors.

\section{Case study: transmission of correlated signals observed
at different nodes}

As discussed in the Introduction, the transmission of correlated
signals observed at different nodes to one or more collectors has
become a topical problem in the recent years, mainly because of the
quick diffusion of Wireless Sensor Networks (WSNs). We consider in
this Section a simple scenario where two independent nodes have to
transmit correlated sensed data to a collector node. Such data,
referred to as $x_i$ and $y_i$, are taken to be i.i.d. correlated
binary randon variables with $P_r\left\{x_i = 1/0\right\} =
P_r\left\{y_i = 1/0\right\} = 0.5$ and correlation $\rho =
P_r\left\{x_i = y_i\right\} > 0.5$. We consider a very simple Joint
Source Channel Decoding (JSCD) technique where no source encoding is
performed (i.e., no compression) but the two transmitters send their
data over independent AWGN channels using the $r = 1/2$ punctured
turbo code described in the previous Section. The independence of
the noise terms in different links is due to the fact that the nodes
are assumed to transmit over orthogonal multiple access channels
(e.g., using frequency division multiple access). At the receiver
two independent decoders performs an iterative decoding scheme
where, at iteration $m$, the first decoder gives an estimation
$x^{(m)}_i$ of $x_i$ and the second decoder gives an estimation
$y^{(m)}_i$ of $y_i$. To achieve this goal, the first/second decoder
observes the signal coming from the first/second channel and
performs turbo decoding taking $y^{(m-1)}_i$/$x^{(m-1)}_i$ as API.
The correlation estimation $\tilde{\rho}$ is evaluated at iteration
$m$ as:
\begin{equation}
    \begin{array}{cl}
    \tilde{\rho}^{(m)} = \frac{1-\sum\limits_{i = 0}^{k-1} x^{(m-1)}_i \oplus y^{(m-1)}_i}{k}
 \label{eq13AA25.6}
 \end{array}
\end{equation}
Note that at first iteration ($m = 0$) neither the correlation nor
the API are available at the two decoders and hence the first
decoding step is performed by setting $\tilde{\rho}^{(0)} = 0.5$. In
this way the decoder does not need any knowledge about the
correlation between the transmitter data.
On the other hand, the theoretical analysis provided in the previous
Sections show that the decoder performance is not very sensitive to
estimation error (see Fig. \ref{Fig8}).
%when
%$\tilde{\rho} < \rho$.
Hence, we expect that the decoder works well even in presence of
imperfect correlation estimation and that it iteratively
converges to achieve an almost perfect correlation estimation.\\
We compare the proposed JSCD technique with the ideal
separation-based strategy where the to-be-transmitted data are
firstly compressed at the minimum achievable compression rate and
then transmitted into the channel by means of turbo channel coding.
Note that in this case the two transmitters must implement
distributed source coding (DSC), and thus they must have a perfect
correlation estimation (supposedly, correlation is still estimated
at the receiver and then it is sent to the transmitters by means of
a feedback channel). On the other hand, even in presence of perfect
correlation estimation, the problem of designing good practical
source codes for correlated sources is still open. Hence, this
second scheme can be considered as an ideal transmission scheme. In
the DSC case, the two sources $x_i$ and $y_i$ are independent (on
account of compression) and, hence, decoding is performed without
any API. To provide a fair comparison with the proposed JSCD
technique we assume that in the separation case the rate of the
channel encoder is lower, so that the global transmission rates is
the same for the two cases. To elaborate, let assume a correlation
$\rho = 0.939$ between the two sources. In this case the joint
entropy of the two information signals is
$H({\mathbf{x}},{\mathbf{y}})$ = $H({\mathbf{x}})$ +
$H({\mathbf{x}}|{\mathbf{y}})$ = $1 - \rho \times log_2(\rho) -
(1-\rho)\times log_2(1-\rho)$ = $1.33$. This means that the two
transmitters may achieve a compression rate of $r_c = 1.33/2 = 2/3$
\footnote{We assume, as usually done for DSC, that the two
transmitters use the same compression rate}. Hence, in order to
achieve the same rate $r=1/2$ as the JSCD case, in the separation
case the channel coding rate may be set to $1/3$. This can be
achieved by using the unpunctured version of the turbo code
described in the previous Section. Moreover, we consider the same
signal-to-noise ratio $SNR = 2r\gamma_b$ for JSCD and DSC, so that
the two schemes are compared for the same overall transmitted rate
and the same same total transmitted energy. Note that, since the
channel rate in the DSC case is 3/2 times lower than in the JSCD
case, the $\gamma_b$ value is $3/2$ times higher (i.e, 1.76 dB
higher). In other terms, we compare the rate $r=1/2$ JSCD scheme
with a given $\gamma_b = \gamma$ dB with the rate $r=1/3$ DSC scheme
with $\gamma_b = \gamma + 1.76$ dB.
 \\
Fig. \ref{Fig9} show a BER comparisons between the JSCD and DSC
scenarios described above. In particular, in  Fig. \ref{Fig9} (a) we
consider an AWGN channel model where the two channels are
characterized by the same $SNR$. In Fig. \ref{Fig9} (b) we instead
consider a block Rayleigh fading channel model \footnote{The fading
is assumed constant over the duration of a frame} where $SNR$ is
exponentially distributed with the same average $E(SNR)$ in the two
channels. As far as the turbo code is of concern, the frame size $k$
of the interleaver is set to $k = 1000$
bits and the maximum number of iterations is set to 10.\\
Note that in the AWGN case, for a target $P_e = 0.00001$, the
performance of the proposed JSDC scheme is only 0.2 dB worse than
the ideal DSC scheme. This assesses the validity of the proposed
iterative JSCD scheme based on turbo coding.  The most interesting
and, dare we say, surprising results is derived in the Rayleigh
case, where the JSDC decoding scheme clearly outperform DSC with a
gain of more then 7 dB for $P_e = 0.001$. The rationale for this
result is that in presence of an unbalanced signal quality from the
two transmitters (e.g., independent fading), leaving a correlation
between the two information signals can be helpful since the better
quality received signal can be used as side information for
detecting the other signal. In other words, the proposed JSCD scheme
allows to get a diversity gain $K$ which is not obtainable by the
DSC scheme. The diversity gain can be measured as the gradient of
the BER curve, which yields $K = 1$ in the DSC case and $K \cong
1.32$ in the JSCD case. Such a diversity gain is due to the inherent
correlation between information signals and, hence, can be exploited
at the receiver without implementing any kind of cooperation between
the transmitters.

\section{Conclusions}

We have derived a novel analysis for evaluating decoding performance
in presence of a-priori information with imperfect correlation
estimation. According to this analysis, it is shown that the
performance depends not only on the codewords' weight, as in
traditional decoding, but also on the information data weight. We
have then validated the proposed analysis in three different
scenarios: convolutional codes, random codes and turbo codes. In
particular, turbo codes have been shown to approach the performance
of infinite length random codes. Moreover, we have proposed an
effective joint source-channel decoding scheme in a wireless sensors
network scenario where two nodes detect correlated sources and
deliver them to a central collector. Experimental results show the
the proposed scheme allows to approach the ideal Slepian-Wolf scheme
in AWGN channel, and to clearly outperform it over fading channels
on account of a diversity gain which can be achieved without
implementing any kind of cooperation between the transmitters.

\newpage
\bibliographystyle{IEEE}

\clearpage \vfill
\vfill

\clearpage
\newpage
\vfill
\vfill

\clearpage
\newpage
\vfill
\vfill

 \clearpage
\newpage
\vfill
\vfill

\clearpage
\newpage
\vfill
\vfill

\clearpage
\newpage
\vfill
\vfill

\clearpage
\newpage
\vfill
\vfill

\clearpage
\newpage
\vfill
\begin{figure}
\begin{center}
\includegraphics[width=1\textwidth]{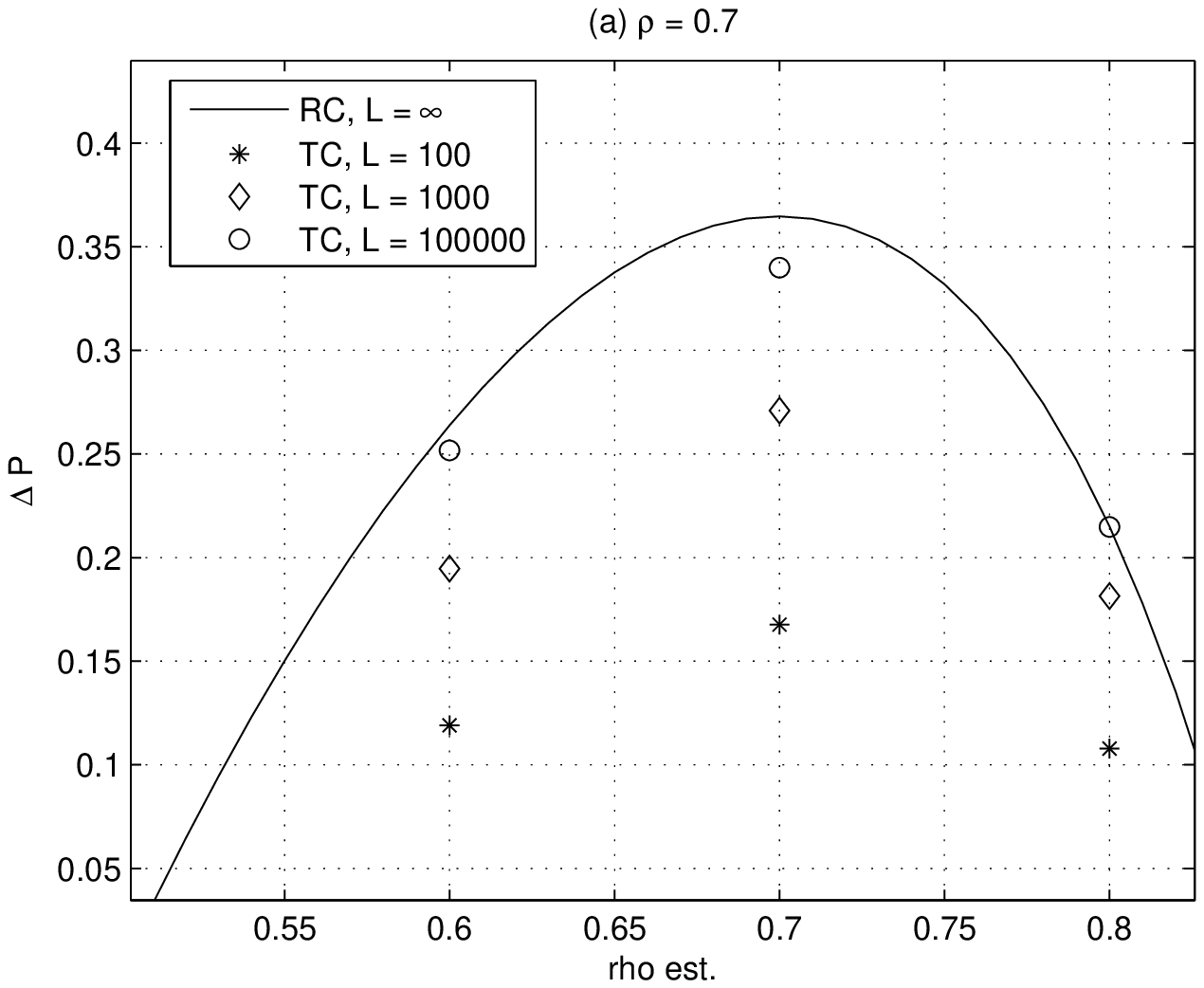}
\end{center}
\end{figure}
\begin{figure}
\begin{center}
\includegraphics[width=1\textwidth]{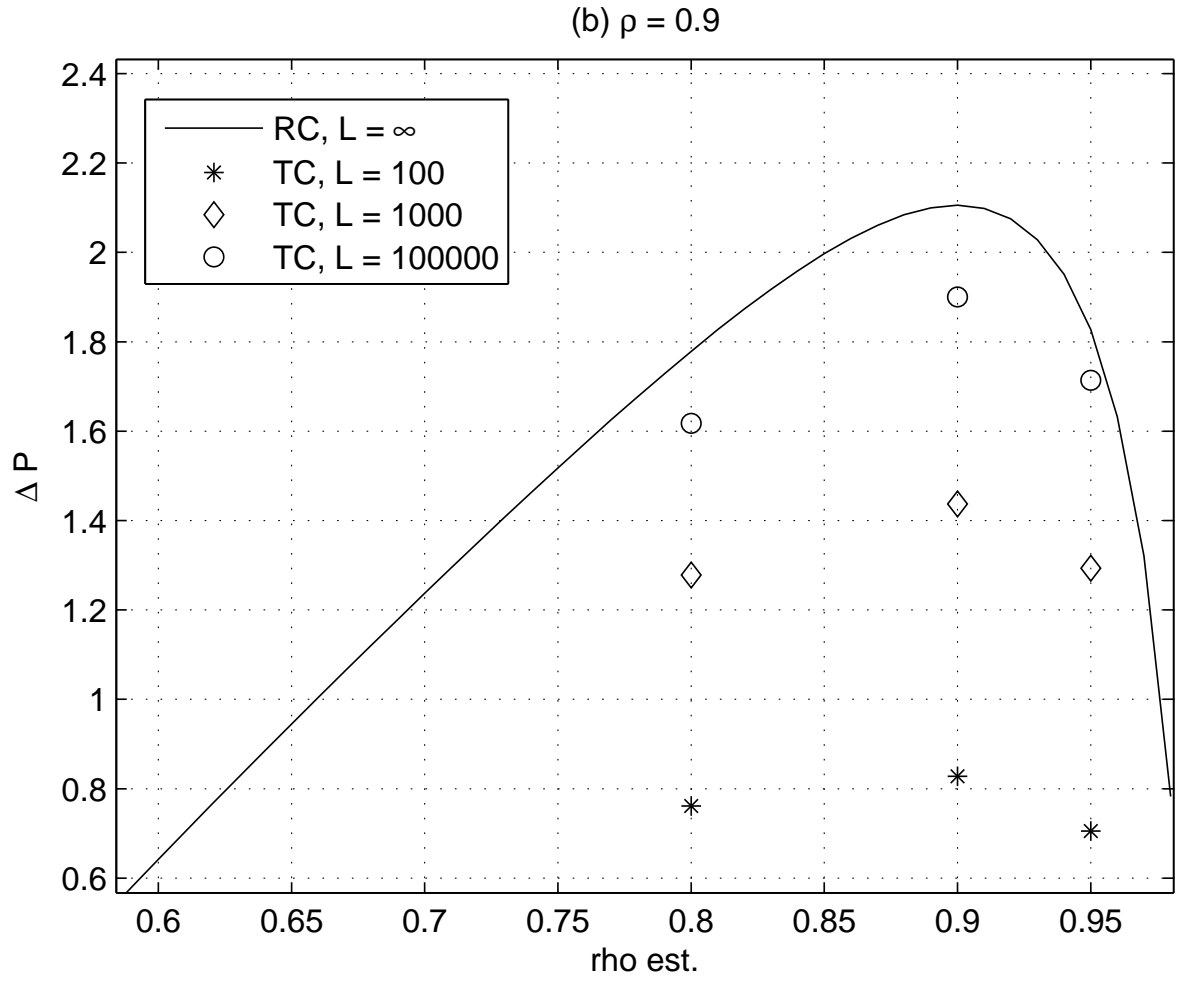}
\end{center}
\caption{$\Delta P$ versus the estimated $\rho$ (i.e.,
${\tilde{\rho}}$ is in the abscissa) for ${P}_{e,r} = 0.00001$:
comparisons between random codes (RC) analysis  with $L = \infty$
(\ref{eq13AA25.1}) and turbo codes ($TC$) with different $L$, for:
(a) $\rho = 0.7$, (b) $\rho = 0.9$}\label{Fig8}
\end{figure}
\vfill

\clearpage
\newpage
\vfill
\begin{figure}
\begin{center}
\includegraphics[width=1\textwidth]{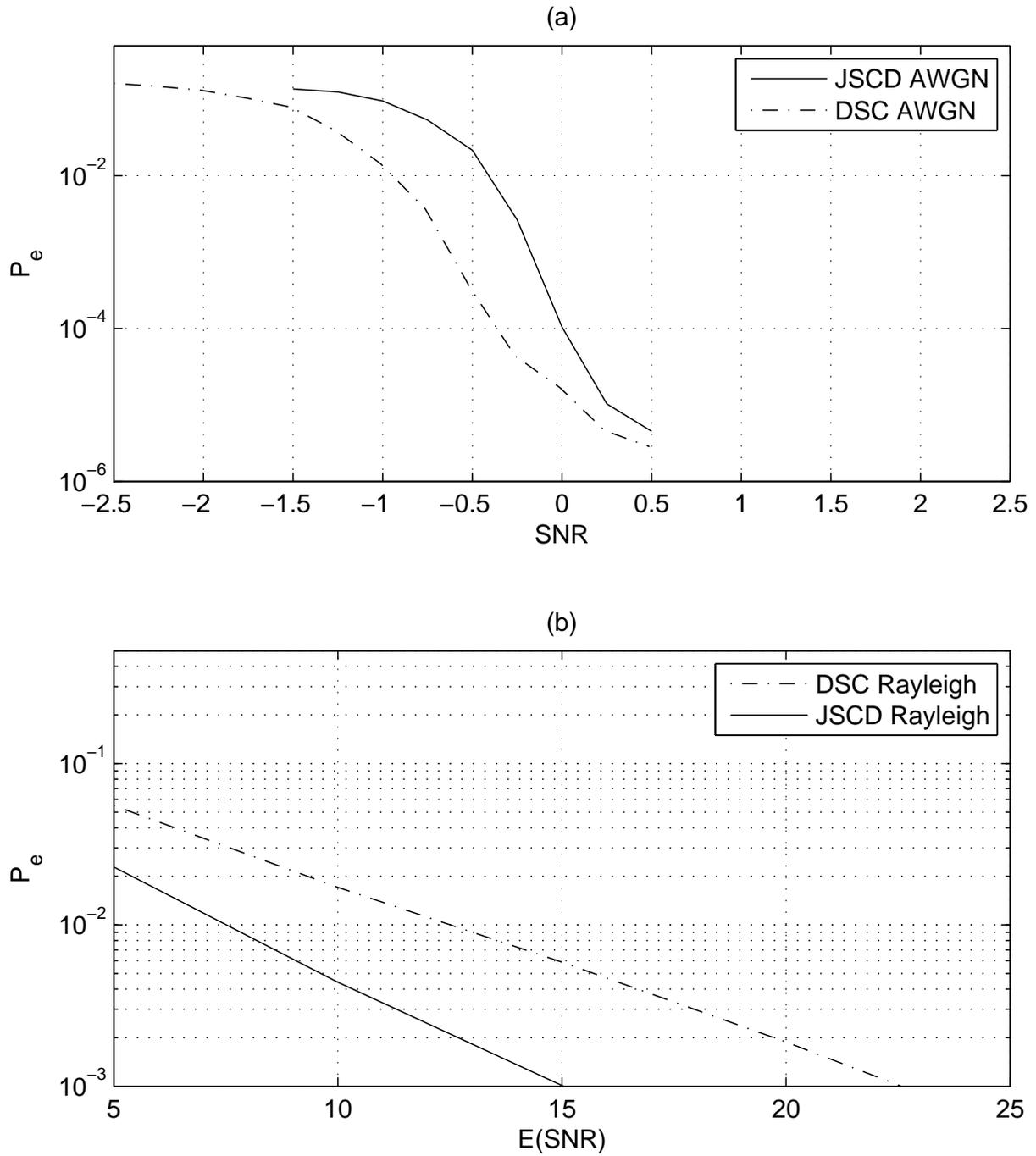}
\end{center}
\caption{BER comparison between JSCD and DSC for : (a) AWGN (b)
Rayleigh fading}\label{Fig9}
\end{figure}
\vfill

\end{document}